\documentclass{article}

\usepackage{euscript}
\usepackage{amssymb}
\usepackage{epsfig}

\newtheorem{lem}{Lemma}
\newenvironment{pf}{{\bf Proof}}{~~{Q.E.D.}}

\newcommand{\ERA}{{\bf ERA}}

\newlength{\opencomm}
\newcommand{\opn}{\hspace*{\opencomm}}
\newlength{\gammalen}
\newcommand{\gammaspc}{\hspace*{\gammalen}}

\newcommand{\DERIV}[2]{\mbox{$\stackrel{\hspace*{-1mm}\mbox{{\tiny#1}}}{\Rightarrow}
                       \hspace*{-1.1em}
                       \raisebox{-1.25mm}{{\tiny #2}}
                       \hspace*{0.4em}$}}

\newcommand{\SEM}[2]{\mbox{$#1[\![#2]\!]$}}
\newcommand{\SOP}[2]{\mbox{$#1\lefteqn{\{}\hspace*{2.7pt}\vert#2\lefteqn{\vert}\}$}}

\newcommand{\CINCL}{\mbox{$\subseteq$}}
\newcommand{\CCUP}{\mbox{$\cup$}}

\newcommand{\ABOT}{\mbox{$\bot^\prime$}}
\newcommand{\ATOP}{\mbox{$\top^\prime$}}
\newcommand{\AINCL}{\mbox{$\supseteq$}}
\newcommand{\ACAP}{\mbox{$\cap$}}
\newcommand{\ACUP}{\mbox{$\cup$}}
\newcommand{\CV}{\mbox{${\EuScript C\EuScript V}$}}
\newcommand{\FP}{\mbox{${\EuScript F\EuScript P}$}}
\newcommand{\TERM}{\mbox{${\EuScript T\EuScript R\EuScript S}$}}

\newcommand{\EQS}{\mbox{${\EuScript E\!\EuScript Q\EuScript S}$}}
\newcommand{\PSEQS}{\mbox{$\wp(\EQS)$}}
\newcommand{\PPSEQS}{\mbox{$\wp(\wp(\EQS))$}}

\newcommand{\ASDE}{\mbox{$\PSEQS(\AINCL,\EQS,\ACAP)$}}

\newcommand{\ACC}[1]{\mbox{${\EuScript A}(#1)$}}

\newcommand{\Co}[1]{\mbox{\bf Co(#1)}}

\newcommand{\DWID}{\mbox{$\lefteqn{\mbox{\raisebox{1.1ex}[0pt][0pt]{\hspace*{0.5pt}$\sim$}}}\bigtriangledown$}}
\newcommand{\MFVS}[1]{\mbox{$#1\backslash_{\mbox{\tiny\bf FVS}}$}}

\newcommand{\OPR}[1]{\mbox{\tt\bf #1}}

\newcounter{numline}
\newcommand{\NM}{\vspace*{-1pt}\stepcounter{numline}\lefteqn{\mbox{\tiny\bf\thenumline:}}}

\newcommand{\Z}{\hspace*{5mm}}

\newcommand{\KW}[1]{{\bf #1}}

\begin{document}

\title{Analysis of Equality Relationships \\ for Imperative Programs}

\author{Pavel Emelianov%
\footnote{This work was partly done when the author was in {\em
Laboratoire d'informatique, Ecole polytechnique}\/ (Palaiseau,
France) and  {\em Ecole normale sup\'erieure d'ingenieur}
(Bourges, France).} \\
Institute of Informatics Systems, \\
6 avenue Lavrentiev \\
630090 Novosibirsk, Russia \\
emelianov@iis.nsk.su}



\maketitle

\begin{abstract}
In this article, we discuss a flow--sensitive analysis of equality
relationships for imperative programs. We describe its semantic
domains, general purpose operations over abstract computational
states (term evaluation and identification, semantic completion,
widening operator, etc.) and semantic transformers corresponding
to program constructs. We summarize our experiences from the last
few years concerning this analysis and give attention to
applications of analysis of automatically generated code. Among
other illustrating examples, we consider a program for which the
analysis diverges without a widening operator and results of
analyzing residual programs produced by some automatic partial
evaluator. An example of analysis of a program generated by this
evaluator is given.

\vskip\baselineskip


\end{abstract}

\section*{Introduction}

{\em Semantic analysis}\/ is a powerful technique for building
effective and reliable programming systems. In
\cite{EmSa/94,Em/96,SAS-4/97} we presented a new kind of a
semantic flow--sensitive analysis designed in the framework of
{\em abstract interpretation}\/ \cite{CC/77,CC/92b,CC/92a}. This
analysis which determines an approximation of sets of invariant
term equalities $t_1=t_2$\/ was called {\em the analysis of
equality relationships for program terms} (hereinafter referred to
as \ERA).

Most traditional static analyses of imperative programs are
interested in finding the (in)equalities of a specific kind
(so--called {\em value analyses}; only they are discussed here)
describing regular approximations (i.e. they have simple
mathematical descriptions and machine representations) of sets of
values: convex polyhedrons/oc\-ta\-hed\-rons/oc\-ta\-gons
\cite{CouHal/78,HalbwachsProyRoumanoff/97,Mine/2001:pado,ClarisoCortadella-SAS04},
affine \cite{Karr/76,GulwaniNecula-POPL03} and congruent
\cite{Gra/91b,Mas/92} hyper-planes, their non-relational
counterparts \cite{CC/77,WZ/91,Gra/89,Mas/93} as well, etc. They
are carefully designed to be reasonable (i.e. they express
non-trivial semantic properties) and effectively computed (i.e.
there are polynomial
algorithms to handle them\footnote{%
An example of this sort is an approximation of value sets by conic
shapes that has only one ``computational disadvantage'': semantic
transformers involve algorithms known to be ${\mathcal NP}$-hard.
Proposed in \cite{BerMar/76} many years ago it did not gain
ground.}) but ``regular'' nature does not allow them to treat well
programs with irregular control/data-flows. Hence of special
interest is investigations of approximations based on sets of
terms which can have potentially arbitrary nature, i.e. they could
be powerful (due to their irregularity) but effectively computed.
One well known example is the set--based analysis
\cite{HeiJaf/94,CC/95}.

In our case, terms represent all expressions computed in programs.
This enables the analysis to take into account different aspects
of program behavior in a unified way. A such unified treatment of
all semantic information allows the analysis to improve its
accuracy. This does not mean that \ERA\/ is a generalization of
all other value analyses (except the {\em constant propagation}
one), because they use different approaches (semantic domains and
transformers) to extract effectively and precisely the limited
classes of semantic properties. In general, the results of the
analyses are not comparable.

\ERA\/ provides interesting possibilities for gathering and
propagating different invariant information about programs in a
unified way. This information can be used both for verification
and optimization purposes. The second is especially interesting
for automatically generated programs: residual, i.e. obtained in
the process of the partial evaluation, and synthesized from
high--level specifications. Due to nature of automatic generation
processes, such programs have specific control flows (for example,
hierarchy of nested conditional statements with specific
conditions; in the case of residual programs this hierarchy is
more deep as ``degree'' of the partial evaluation increases) that
can be successfully optimized on base of gathered invariant
information.

Besides the peculiarity of \ERA\/ mentioned above, let us discuss
some common properties of the semantic analyses. Such taxonomic
properties of the analysis algorithms as the attribute
(in)dependence, context (in)sensitivity, flow (in)sensitivity,
scalability and some other properties are well known. However, it
is the author's opinion that a notion of ``{\em interpretability
of a semantic analysis}''\/ has not been considered adequately
yet. Here the interpretability of analysis means how extensively
the properties of primitive operations of the language
(arithmetical, logical, etc.) and type information are allowed for
analyzing and can be handled when the analysis works.

One extreme point of view on the interpretability is an approach
accepted in the ``pure'' program scheme theory where no
interpretations of functional symbols or type information are allowed\footnote{%
More precisely, there exist some works in the program scheme
theory where some semantic interpretations of functional symbols
(like commutativity of superposition $f\circ g\equiv g\circ f$,
etc.) are considered.}. Unfortunately, the results obtained under
this approach are not reasonably strong. Nevertheless, it must be
underscored that \ERA\/ dates back to V. Sabelfeld's works in the
program scheme theory \cite{Sa/79,Sa/80}. Another extreme leads to
the complete description of the program behavior that is also not
workable. Obviously, it is closely allied to its flow sensitivity
(ignoring some part of semantic information does not allow us to
treat precisely some control flow constructs of analyzed programs)
and its scalability (attempts to take into account large quantity
of semantic properties, for example, using some theorem prover
which is invoked while an semantic analyzer works and deduces new
properties, can lead to combinatorial explosion for abstract
computational states).

It is possible that the interpretability has not been highlighted
enough, because most of analysis algorithms take into account the
limited classes of primitive operations and type information and
they cannot be enriched in some natural way. For example, an
interval analysis is not able to incorporate congruence properties
in some natural way, etc\footnote{Of course, it is possible to use
sophisticated approaches for combinations of analyses but it
introduces complicated problems under implementations and it is
not an enrichment of original ones.}. Essentially another case is
\ERA\/ where we have a choice to handle expressiveness of the
analysis. We intend to illustrate the notion of ``interpretability
of analysis'', its importance and usefulness on this example of
analysis.

Among the analyses closely related to ours we would like to point
out the following. A semantic analysis for detection of equalities
between program variables (and simple relationships among them)
was described in \cite{AWZ/88}. It makes a list of sets of
variables discovered to be equal by using the Hopcroft's
partitioning algorithm for finite--state automata. This algorithm
being quite efficient is not however precise enough. Further
value-numbering techniques were developed in
\cite{RuthingKnoopSteffen-SAS99,Gargi-PLDI02,GulwaniNecula-SAS04}.
These algorithms demonstrate that adequacy of value numbering is
resource-consuming. For example, in the last case time complexity
of the algorithm is $O(k^3jN)$\/ where $k$\/ is a number of
program variables, $j$\/ is a number join-points, and $N$\/ is
program size.

Another important example is the set-based analysis
\cite{HeiJaf/94} mentioned above. Here approximating sets of terms
are found with resolving some system of set-theoretical equations.
Formal grammars were used for an analysis of recursive data
structures of functional languages (see, for example,
\cite{Jo/87}). Formal languages were applied to coding of memory
access paths in \cite{Deu/94,Venet/99} and values of program
variables in the set-based analysis. \cite{CC/95} established
common foundations connecting and generalizing different
approaches using formal languages to represent semantic
properties. Of course, we should mention techniques from the
automatic proof theory and the term rewriting theory which can be
widely applied both at the analysis stage to improve its accuracy
and the post-processing stage to present its results to the user.

This article is organized as follows: In {\bf Sections
\ref{PConcrete}} and {\bf\ref{PAbstract}} we describe the semantic
properties, concrete and abstract, respectively, which are
considered in \ERA. In {\bf Section \ref{OpBasic}} we discuss some
basic operations over the semantic properties used to define the
semantic transformers which are next presented in {\bf Section
\ref{STransformers}}. In {\bf Section \ref{WideningOp}} we
consider a widening operator and  the complexity of \ERA\/ is
discussed. Finally, {\bf Section \ref{Experiments}} describes
processing of \ERA\/ invariants and presents some results of our
experiments with \ERA. In {\bf Appendix} an example of analysis of
some residual program is considered.

\section{Properties of interest}

\subsection{Concrete properties}\label{PConcrete}

A usual choice for the description of the operational semantics is
a specification of some transition relationship on the pairs
$<${\em control point, state of program memory}$>$ (see, for
example, \cite{PFA/81,HOOTS/98}) where the states of program
memory are described by mapping the cells of memory into a
universe of values. Here variables (groups of cells) and their
values (constants) are in asymmetric roles. Another example of
``asymmetry'': manipulations over the structured objects of
programs (arrays, records etc.) are not so transparent as over the
primary ones.  To describe the operational semantics for \ERA, we
used another approach. All objects of a program are considered to
be ``identical'' in the following meaning.

Let \CV\/ be a set of $0$--ary symbols representing variables and
constants. The last ones may be of the following kinds: scalars,
compositions over scalars (i.e., constant arrays, records, etc.),
names of record fields, and {\em indefiniteness}. Let \FP\/ be a
set of n--ary ({\em functional}) symbols which represent primitive
operations of programming languages: arithmetic, logic, type
casting, and all the kinds of memory addressing, as well. Let
\TERM\/ be a set of well-formed terms over \CV\/ and \FP,
hereinafter referred to as {\em program terms}. They represent
expressions computed during execution of a program. So, as a state
of program memory we take a reflexive, symmetrical, and transitive
relationship (i.e., {\em equivalence relationship}) over \TERM.
The relationship defines some set of term equalities which we use
to describe the operational semantics and call a {\em computation
state}.
%
%

\begin{table}
\begin{center}
\fbox{\fontsize{8}{10}\selectfont
\begin{tabular}{lcc}
&&\\
       & {\sc then}-branch   &     {\sc else}-branch\\
&&\\
\hline
&&\\
\sc entry
       & $\!\!\!\!\left\{\rule[10mm]{0pt}{0pt}\!\!\left\{\!\!\mbox{\begin{tabular}{l}
                       a[1]=1,a[2]=j=2,a[3]=i=3,\\
                       a=\fbox{1}\fbox{2}\fbox{3},ODD(x)=TRUE\\
          \end{tabular}}\!\!\right\}\!\!\right\}$
       & $\left\{\rule[10mm]{0pt}{0pt}\!\!\left\{\!\!\mbox{\begin{tabular}{l}
                       a[1]=1,a[2]=j=2,a[3]=i=3,\\
                       a=\fbox{1}\fbox{2}\fbox{3},ODD(x)=FALSE\\
          \end{tabular}}\!\!\right\}\!\!\right\}$
       \\
&&\\
\sc exit
       & $\!\!\!\!\left\{\rule[10mm]{0pt}{0pt}\!\!\left\{\!\!\mbox{\begin{tabular}{l}
                       a[1]=i=j=1,a[2]=2,a[3]=3,\\
                       a=\fbox{1}\fbox{2}\fbox{3},ODD(x)=TRUE\\
          \end{tabular}}\!\!\right\}\!\!\right\}$
       & $\left\{\rule[10mm]{0pt}{0pt}\!\!\left\{\!\!\mbox{\begin{tabular}{l}
                       a[1]=i=j=3,a[2]=2,a[3]=1,\\
                       a=\fbox{3}\fbox{2}\fbox{1},ODD(x)=FALSE\\
          \end{tabular}}\!\!\right\}\!\!\right\}$
       \\
&&\\
\hline
&&\\
& \hfill \sc exit ~of~ $\lefteqn{\mbox{\footnotesize\sc if--statement}}$ & \\
&&\\
{\Large$\pi=$}
       & $\!\!\!\!\left\{~\rule[10mm]{0pt}{0pt}\!\!\left\{\!\!\mbox{\begin{tabular}{l}
                       a[1]=i=j=1,a[2]=2,a[3]=3,\\
                       a=\fbox{1}\fbox{2}\fbox{3},ODD(x)=TRUE
                       \smash{\lefteqn{\mbox{\hspace*{3mm}\huge,}}}\\
          \end{tabular}}\!\!\right\}\!\!\right.$
       & \hspace*{-2mm}$\!\!\left.\!\!\left\{\mbox{\begin{tabular}{l}
                       a[1]=i=j=3,a[2]=2,a[3]=1,\\
                       a=\fbox{3}\fbox{2}\fbox{1},ODD(x)=FALSE\\
          \end{tabular}}\!\!\right\}\rule[10mm]{0pt}{0pt}~\!\!\right\}$
       \\
&&\\
\end{tabular}
}
\end{center}
\caption{Description of collecting semantics for {\bf Example 1}
         (here the constant \fbox{$c_1$}\fbox{$c_2$}\fbox{$c_3$}
         represents constant arrays).}\label{Collecting}
\end{table}

Suppose that the following code

\begin{tabbing}
\quad \=\quad \=\quad           \kill
   \KW{var} x,i,j : \KW{integer}; a : \KW{array} [1..3] \KW{of integer}=\{1,2,3\};\\
   ...\\
   i := 3;\\
   j := i-1;\\
   \KW{if} \KW{odd}(x) \KW{then}\\
\>   i := i \KW{mod} j;\\
\>   j := 1;\\
   \KW{else}\\
\>   j := a[i];\\
\>   a[i] := a[1];\\
\>   a[1] := j\\
   \KW{end}\\
   ...
\end{tabbing}
\begin{center}{\bf Example 1}\end{center}

\noindent is executed at least twice for the different parities of
the variable {\sc x}. In {\bf Table \ref{Collecting}} static
semantics for five control points is given. We present a minimum%
\footnote{A set of equalities can be completed with any number of
consistent equalities.} subset of term equalities concerning
dynamic behavior of the piece. We shall use the property
{\large$\pi$}\/ (see {\bf Table \ref{Collecting}}) to illustrate
our further reasoning.

Formally it is described as follows. Let \EQS\/ be a set of all
equalities of the terms from \TERM, i.e.,
$\EQS=\{t_1=t_2~\vert~t_1, t_2\in\TERM\}$. A set $S\in\PSEQS$\/ is
a computation state interpreted in the following way. For each
equality $t_i=t_j\in S$\/ values of the expressions represented by
$t_i$\/ and $t_j$\/ must be equal at this point for some execution
trace. We take the set \PPSEQS\/ as a set for a concrete semantic
domain describing {\em the collecting semantics}\/ of \ERA. So, an
element of the concrete semantic domain is a set of computation
states. For a particular point in a particular program it is a set
of computation states each of them corresponds to some execution
trace in the program that reaches this point.

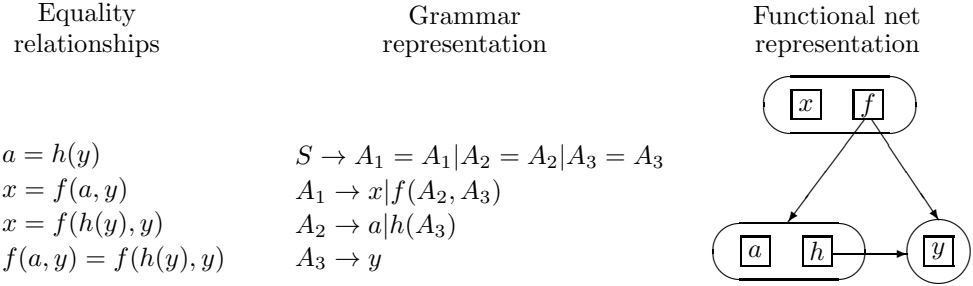
\begin{figure}
\centering
\unitlength=0.75mm
\linethickness{0.4pt}
\begin{picture}(170,61.50)
\put(15.00,52.50){\makebox(0,0)[cc]{Equality}}
\put(15.00,47.00){\makebox(0,0)[cc]{relationships}}
\put(82.00,52.50){\makebox(0,0)[cc]{Grammar}}
\put(82.00,47.00){\makebox(0,0)[cc]{representation}}
\put(148.00,52.50){\makebox(0,0)[cc]{Functional net}}
\put(148.00,47.00){\makebox(0,0)[cc]{representation}}
\put(0,8){\shortstack[l]{$a=h(y)$\\$x=f(a,y)$\\
                          $x=f(h(y),y)$\\$f(a,y)=f(h(y),y)$}}
\put(52,8){\shortstack[l]{$S\rightarrow A_1=A_1|A_2=A_2|A_3=A_3$\\
                           $A_1\rightarrow x|f(A_2,A_3)$\\
                           $A_2\rightarrow a|h(A_3)$\\
                           $A_3\rightarrow y$}}
\put(148.50,36.50){\oval(27.00,10.00)[]}
\put(140.00,34.00){\framebox(5.00,5.00)[cc]{$x$}}
\put(151.00,34.00){\framebox(5.00,5.00)[cc]{$f$}}
\put(153.00,34.00){\vector(-3,-4){13.67}}
\put(139.50,10.50){\oval(27.00,10.00)[]}
\put(131.00,8.00){\framebox(5.00,5.00)[cc]{$a$}}
\put(142.00,8.00){\framebox(5.00,5.00)[cc]{$h$}}
\put(166.00,10.50){\circle{11.0}}
\put(163.50,8.00){\framebox(5.00,5.00)[cc]{$y$}}
\put(154.00,34.00){\vector(2,-3){12.0}}
\put(147.00,10.00){\vector(1,0){13.50}}
\end{picture}
\caption{Semantic properties and their representations. The set of equalities
does not contain trivial equalities like $x=x$\/ and equalities given by
the symmetry of the equality relationship.}\label{SemPrps}
\end{figure}

Properties considered in \ERA\/ are presented by means of
context--free grammars  ${\bf G}=({\EuScript N},{\EuScript
T},{\EuScript P},S)$\/ where ${\EuScript N}$\/ is a finite set of
nonterminals denoted by capital letters, $S\in{\EuScript N}$\/ is
the initial symbol of the grammar ${\bf G}$, ${\EuScript
T}=\CV\cup\FP\cup\{~(,~),~=,$~\raisebox{1mm}{{\Huge ,}}$\}$\/ is a
finite set of terminal symbols, and ${\EuScript P}$\/ is a finite
set of grammar rules. We do not give their precise description
because we shall use quite simple machinery from formal languages
theory that is not an object of considerations itself and serves
for demonstrations. We could use functional nets language as well
but it is rather machine--oriented and is not widely used. We
expect that all these descriptive ways become apparent from the
examples on {\bf Figure \ref{SemPrps}} and further ones.

In that way, a state of computation is represented by a language
$L{\bf(G)}$\/ generated by some grammar {\bf G} of the described
form. If for $A\in{\EuScript N} ~:~ \linebreak A\DERIV{+}{G}t_1
~\wedge~ A\DERIV{+}{G}t_2$, i.e. $t_{1}=t_{2}\in L$({\bf G}), we
say that the nonterminal $A$\/ and the language $L$({\bf G}) know
the terms $t_{1},~t_{2}$. Obviously, such a grammar representation
has some superfluous ``syntactic sugaring''. We can use
$S\rightarrow A$\/ rules only and say about $A$--nonterminals as
classes of equal values.

Evidently we may suppose that the set of rules ${\EuScript P}$\/
does not contain rules having identical right parts. It is
convenient to consider the grammars which do not contain useless
and redundant nonterminals and rules. A rule is useless if it
produces only one term (the language knows only a trivial equality
like $t=t$) and this term is not an argument of other terms. A
nonterminal is useless if it does not participate in derivations
of sentential forms \cite{AU/72}. Such grammars can arise as
result of operations on grammars. If nonterminals and rules are
useless or redundant it is possible to remove them (see {\bf Lemma
\ref{DeleteTerms}}). This operation called {\bf state reduction}
consist in detecting and removing a set of useless/redundant
nonterminals/rules. They are revealed with the well--known
incremental markup algorithms (see, for example, \cite{AU/72}).

For illustrating purposes we shall use special functional nets for
these grammars. Here nonterminals are represented by ovals
containing $0$--ary and functional symbols from right parts of
rules. Arcs from functional symbols to ovals represent argument
dependencies ordered from left to right (see {\bf Figure
\ref{SemPrps}}).

\subsection{Abstract properties}\label{PAbstract}

It is an interesting peculiarity of \ERA\/ that abstract (i.e.
approximate) properties have the same nature as the computation
states of the operational semantics. Formally this approximation
is defined as follows.

\subsubsection{Functions of abstraction and concretization}

Given a concrete property $\pi\in\PPSEQS$\/ and an abstract
property $\tilde{\pi}\in\PSEQS$, the abstraction function
$\alpha:\PPSEQS\rightarrow\PSEQS$\/ and the concretization one
$\gamma:\PSEQS\rightarrow\PPSEQS$\/ are defined in the following
way
\[
\begin{array}{lcl}
\alpha(\pi)~=~\left\{
\begin{array}{lcl}
\EQS,              & ~ & \mbox{if } \pi = \emptyset,\\
\mbox{$\displaystyle\bigcap_{S\in\pi}$}S
                   &   & \mbox{otherwise}
\end{array}
\right.& \mbox{~~and~~} &
\gamma(\tilde{\pi}) = \CCUP\{~\pi~\vert~\alpha(\pi)~\AINCL~\tilde{\pi}~\}\\
\end{array}
\]
where \CCUP\/ is the set--theoretical union on \PPSEQS\/ and
\AINCL\/ and \ACAP\/ are the set--theoretical inclusion and
intersection of the languages (i.e. on \PSEQS) respectively. We
take the empty language as infimum \ABOT\/ (there are no computed
expressions) of the semi-lattice of abstract semantic properties.
The supremum \ATOP\/ (an inaccessible computation state) is the
language containing all possible equalities of program terms. Also
\CINCL\/ is the set--theoretical inclusion on \PPSEQS\/

\begin{lem}
The abstraction function $\alpha$\/ is monotonic.
\end{lem}

\begin{pf}
The function $\alpha$\/ is monotonic iff $\forall
\pi_1,\pi_2\in\PPSEQS : \pi_1\CINCL\pi_2 \Rightarrow \linebreak
\alpha(\pi_1)\AINCL\alpha(\pi_2)$. Because $\pi_1\CINCL\pi_2$,
then $\pi_2=\pi_1\CCUP(\pi_2\setminus\pi_1)$. So, we have
$\alpha(\pi_1)\AINCL \linebreak
\alpha(\pi_1\CCUP(\pi_2\setminus\pi_1))=
\alpha(\pi_1)\cap\alpha(\pi_2\setminus\pi_1)=\alpha(\pi_1)\cap(\cap_{S\in\pi_2\setminus\pi_1}S)$.
\end{pf}

\noindent $\gamma(\tilde{\pi})$\/ is the most imprecise element of \PPSEQS\/
that can be soundly approximated by $\tilde{\pi}\in\PSEQS$.
For the example of {\bf Table \ref{Collecting}}, the best approximation
of the concrete property {\large$\pi$} is
\[
\alpha(\pi)=\left\{\rule{0pt}{5mm}a[1]=i=j,a[2]=2\right\}.~~~~~~~~~~~\mbox{{\bf$(*)$}}
\]

%
%

\subsubsection{Intersection of \ERA--languages}

\begin{figure}[t]
{\small
\centering
\unitlength=0.70mm
\linethickness{0.4pt}
\begin{picture}(150.00,103.00)
\put(26.50,95.50){\oval(25.00,9.00)[]}
\put(19.00,93.00){\framebox(5.00,5.00)[cc]{$f$}}
\put(29.00,93.00){\framebox(5.00,5.00)[cc]{$g$}}
\put(11.50,70.50){\oval(25.00,9.00)[]}
\put(4.00,68.00){\framebox(5.00,5.00)[cc]{$z$}}
\put(14.00,68.00){\framebox(5.00,5.00)[cc]{$f$}}
\put(34.00,68.00){\framebox(5.00,5.00)[cc]{$x$}}
\put(44.00,68.00){\framebox(5.00,5.00)[cc]{$a$}}
\put(54.00,68.00){\framebox(5.00,5.00)[cc]{$y$}}
\put(46.50,70.50){\oval(35.00,9.00)[]}
\put(30.00,93.00){\vector(-2,-3){12.00}}
\put(21.00,93.00){\vector(-2,-3){12.00}}
\put(33.00,93.00){\vector(3,-4){13.67}}
\put(32.00,61.50){\oval(30.00,3.00)[b]}
\put(17.00,68.00){\line(0,-1){7.00}}
\put(47.00,61.00){\vector(0,1){5.00}}
\put(103.50,95.50){\oval(25.00,9.00)[]}
\put(96.00,93.00){\framebox(5.00,5.00)[cc]{$f$}}
\put(106.00,93.00){\framebox(5.00,5.00)[cc]{$g$}}
\put(88.50,70.50){\oval(25.00,9.00)[]}
\put(81.00,68.00){\framebox(5.00,5.00)[cc]{$x$}}
\put(91.00,68.00){\framebox(5.00,5.00)[cc]{$f$}}
\put(111.00,68.00){\framebox(5.00,5.00)[cc]{$f$}}
\put(121.00,68.00){\framebox(5.00,5.00)[cc]{$a$}}
\put(131.00,68.00){\framebox(5.00,5.00)[cc]{$y$}}
\put(107.00,93.00){\vector(-2,-3){12.00}}
\put(98.00,93.00){\vector(-2,-3){12.00}}
\put(110.00,93.00){\vector(3,-4){13.50}}
\put(141.00,68.00){\framebox(5.00,5.00)[cc]{$z$}}
\put(128.50,70.50){\oval(45.00,9.00)[]}
\put(111.00,70.00){\vector(-1,0){10.00}}
\put(111.50,61.50){\oval(35.00,3.00)[b]}
\put(94.00,68.00){\line(0,-1){7.00}}
\put(129.00,61.00){\vector(0,1){5.00}}
\put(129.00,86.00){\makebox(0,0)[cc]{$L_{2}$}}
\put(51.00,86.00){\makebox(0,0)[cc]{$L_{1}$}}
\put(44.50,47.50){\oval(25.00,9.00)[]}
\put(37.00,45.00){\framebox(5.00,5.00)[cc]{$f$}}
\put(47.00,45.00){\framebox(5.00,5.00)[cc]{$g$}}
\put(48.00,45.00){\vector(-2,-3){12.00}}
\put(39.00,45.00){\vector(-2,-3){12.00}}
\put(51.00,45.00){\vector(3,-4){13.67}}
\put(29.00,20.00){\framebox(5.00,5.00)[cc]{$f$}}
\put(31.50,22.50){\oval(19.00,9.00)[]}
\put(64.50,22.50){\oval(25.00,9.00)[]}
\put(57.00,20.00){\framebox(5.00,5.00)[cc]{$a$}}
\put(67.00,20.00){\framebox(5.00,5.00)[cc]{$y$}}
\put(48.50,13.50){\oval(33.00,3.00)[b]}
\put(65.00,13.00){\vector(0,1){5.00}}
\put(32.00,20.00){\line(0,-1){7.00}}
\put(98.50,47.50){\oval(25.00,9.00)[]}
\put(91.00,45.00){\framebox(5.00,5.00)[cc]{$f$}}
\put(101.00,45.00){\framebox(5.00,5.00)[cc]{$z$}}
\put(96.00,20.00){\framebox(5.00,5.00)[cc]{$x$}}
\put(98.50,22.50){\oval(19.00,9.00)[]}
\put(94.00,45.00){\vector(1,-4){4.50}}
\put(133.00,36.00){\makebox(0,0)[cc]{$L_{1}\ACAP L_{2}$}}
\end{picture}
\caption{Intersection of computation states.}\label{Intersection}
}
\end{figure}
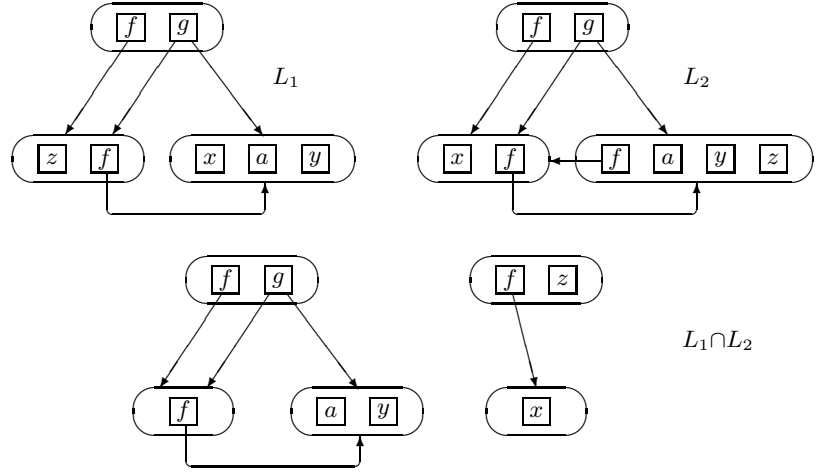

Finding of the intersection of context--free languages is an
undecidable problem in the general case. In our case, for the
languages of term equalities, an algorithm exists (see an example
at {\bf Fig. \ref{Intersection}}). It is similar to constructing a
Cartesian product of automata.

\vskip 0.75\baselineskip
\noindent{\bf Algorithm.} Intersection of two languages of term equalities.\\
Input:  grammars  ${\bf G_1}=({\EuScript N}_1,{\EuScript T},{\EuScript P}_1,S_1)$
        and ${\bf G_2}=({\EuScript N}_2,{\EuScript T},{\EuScript P}_2,S_2)$.\\
Output: grammar ${\bf G}=({\EuScript N},{\EuScript T},{\EuScript
P},S)$ such that
        $L({\bf G})=L({\bf G_1})\cap L({\bf G_2})$.\\
Description:
\begin{enumerate}
\item Let ${\EuScript N}=\{ \langle N_1,N_2\rangle \vert%
          N_1\in{\EuScript N}_1~\&~N_2\in{\EuScript N}_2 \}={\EuScript N}_1\times{\EuScript N}_2$.

\item The set of rules ${\EuScript P}$ is defined as follows:
      \begin{itemize}
      \item The rule $\langle N_1,N_2\rangle\rightarrow t$ is introduced
            if and only if
            $t\in\CV~~\&~~N_1\rightarrow t\in{\EuScript P}_1~~\&~~
                          N_2\rightarrow t\in{\EuScript P}_2$.
      \item The rule $\langle N_1,N_2\rangle\rightarrow t(%
            \langle N_1^1,N_2^1\rangle,\ldots,\langle N_1^k,N_2^k\rangle)$
            is introduced if and only if
            $t\in\FP~~\&~~N_1\rightarrow t(N_1^1,\ldots,N_1^k)\in{\EuScript P}_1
                    ~~\&~~N_2\rightarrow t(N_2^1,\ldots,N_2^k)\in{\EuScript P}_2$.
      \end{itemize}

\item Add rules $S\rightarrow\langle N_1,N_2\rangle=\langle N_1,N_2\rangle$
      for the initial nonterminal $S$ of {\bf G} and for all
      $N_1\in{\EuScript N}_1\setminus\{S_1\}, N_2\in{\EuScript N}_2\setminus\{S_2\}$.

\item Apply {\em state reduction}.
\end{enumerate}

The described algorithm of intersection is very ``na\"{\i}ve'' and
impractical. To improve it we should choose a more efficient
strategy for generating functional symbols. To do this we first do
a topological sorting of the functional symbols appearing in the
right parts of rules; intersect the $0$--ary symbol sets; and then
generate the next functional symbol in conformity with its
topological order if and only if all arguments of this symbol
already exist in the new grammar. For practical cases this
intersection can be done in an linear average time with respect to
grammar size and linear space. It demands quadratic time in the
worst case.

\subsection{Operations over semantic properties}\label{OpBasic}

Now we shall discuss some basic operations over semantic properties \linebreak
$\SOP{}{\centerdot}:\PSEQS\rightarrow\PSEQS$\/ used to define the semantic
transformers of \ERA. For all \SOP{}{\centerdot}~ the notation
\SOP{\SOP{L}{\centerdot}}{\centerdot} means
\SOP{(\SOP{L}{\centerdot})}{\centerdot}.

\subsubsection{Removing terms}

Operations over abstract computation states use certain common
transformation of the sets of term equalities which consists in {\bf removing}
some subset $L^\prime$. The following statement holds.

\begin{lem}\label{DeleteTerms}
Removing any subset of term equalities preserves  correctness of an
approximation.
\end{lem}

\begin{pf}
It easy to see that
\[
\begin{array}{lcl}
\CCUP\{~\pi~\vert~\alpha(\pi)~\AINCL~\tilde{\pi}~\}=\gamma(\tilde{\pi})
&~\CINCL~&\gamma(\tilde{\pi}\setminus L^\prime)=\CCUP\{~\pi~\vert~\alpha(\pi)~\AINCL~(\tilde{\pi}\setminus L)~\}=\\
&&\gammaspc\CCUP\{~\pi~\vert~\mbox{$\displaystyle\bigcap_{L\in\pi}$}(L\ACUP
                             L^\prime)~\AINCL~\tilde{\pi}~\},
\end{array}
\]
which states that removing term equalities makes the approximation more rough
but it does preserve its correctness.
\end{pf}

\noindent For (${\fontsize{17}{17}\selectfont*}$), for example,
$\gamma(\{\rule{0pt}{5mm}a[1]=i=j,a[2]=2\})\CINCL \gamma(\{a[1]=i\})$.

We shall write \SOP{L}{\downarrow t} and \SOP{L}{\downarrow T} for
single term and term set removing followed by the state reduction
operation defined above.

\subsubsection{Term evaluation}\label{Term-evaluation}

We define the abstract semantics for an {\bf evaluation of
a term} in an abstract computation state. The result of the
term evaluation is a state knowing the evaluated term.

\vskip\baselineskip
\noindent{\bf Term evaluation} \SOP{L}{t}.

\begin{enumerate}
\item If $t=t\in L$, then $\SOP{L}{t}=L$.
\item Otherwise, if $t\in\CV$, then add the new rules
      $S\rightarrow A=A$ and $A\rightarrow t$\/ to the
      grammar ${\bf G}$, where $A$\/ is a nonterminal
      which does not exist in ${\bf G}$.
\item Otherwise, if $t=f(t_{1},\ldots,t_{n})$\/ where
      $f\in\FP$\/ is a functional $n$--ary symbol and
      the sub-terms $t_{1},\ldots,t_{n}$\/ have been calculated
      (i.e. there exist derivations
      $A_1\DERIV{+}{G}t_1,\ldots,A_n\DERIV{+}{G}t_n$),
      then add the new rules $S\rightarrow A=A$\/ and
      $A\rightarrow f(A_1,\ldots,A_n)$\/ to the grammar
      ${\bf G}$, where $A$\/ is a nonterminal which
      still does not exist in ${\bf G}$.
\end{enumerate}
To improve an accuracy of analysis we can take into account, for example, the
commutativity of primitive operations.
If $L$\/ knows $f(t_1,t_2)$\/ and $f$\/ is commutative then $\SOP{L}{f(t_2,t_1)}=L$.

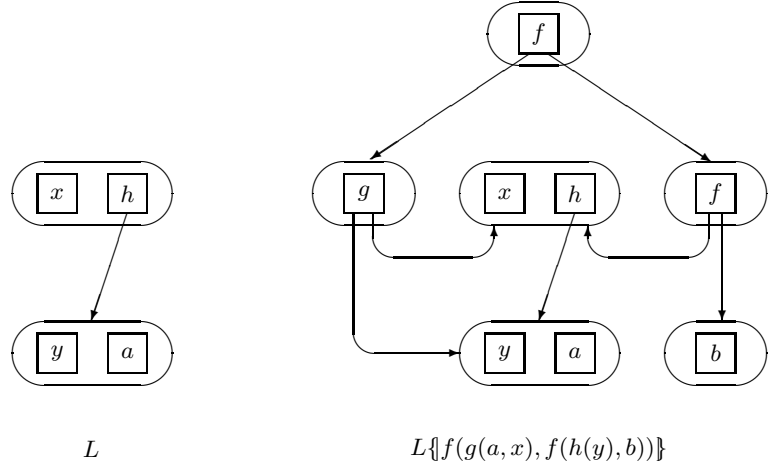
\begin{figure}[t]
\centering
{\small
\unitlength=0.85mm
\linethickness{0.4pt}
\begin{picture}(118.00,70.00)
\put(12.50,15.00){\oval(25.00,10.00)[]}
\put(4.00,12.00){\framebox(6.00,6.00)[cc]{$y$}}
\put(15.00,12.00){\framebox(6.00,6.00)[cc]{$a$}}
\put(12.50,40.00){\oval(25.00,10.00)[]}
\put(4.00,37.00){\framebox(6.00,6.00)[cc]{$x$}}
\put(15.00,37.00){\framebox(6.00,6.00)[cc]{$h$}}
\put(18.00,37.00){\vector(-1,-3){5.67}}
\put(82.50,15.00){\oval(25.00,10.00)[]}
\put(74.00,12.00){\framebox(6.00,6.00)[cc]{$y$}}
\put(85.00,12.00){\framebox(6.00,6.00)[cc]{$a$}}
\put(82.50,40.00){\oval(25.00,10.00)[]}
\put(74.00,37.00){\framebox(6.00,6.00)[cc]{$x$}}
\put(85.00,37.00){\framebox(6.00,6.00)[cc]{$h$}}
\put(88.00,37.00){\vector(-1,-3){5.67}}
\put(79.33,62.00){\framebox(6.00,6.00)[cc]{$f$}}
\put(52.00,37.00){\framebox(6.00,6.00)[cc]{$g$}}
\put(107.00,37.00){\framebox(6.00,6.00)[cc]{$f$}}
\put(107.00,12.00){\framebox(6.00,6.00)[cc]{$b$}}
\put(81.00,62.00){\vector(-3,-2){25.00}}
\put(83.67,62.00){\vector(3,-2){25.67}}
\put(99.50,33.33){\oval(19.00,6.67)[b]}
\put(109.00,37.00){\line(0,-1){4.00}}
\put(90.00,33.00){\vector(0,1){2.00}}
\put(111.00,37.00){\vector(0,-1){17.00}}
\put(65.83,33.33){\oval(19.00,6.67)[b]}
\put(56.33,37.00){\line(0,-1){4.00}}
\put(75.33,33.00){\vector(0,1){2.00}}
\put(57.17,18.33){\oval(7.67,6.67)[lb]}
\put(57.00,15.00){\vector(1,0){13.00}}
\put(53.33,37.00){\line(0,-1){19.00}}
\put(12.33,0.00){\makebox(0,0)[cc]{$L$}}
\put(82.33,0.00){\makebox(0,0)[cc]{$\SOP{L}{f(g(a,x),f(h(y),b))}$}}
\put(110.00,15.00){\oval(16.00,10.00)[]}
\put(110.00,40.00){\oval(16.00,10.00)[]}
\put(82.33,65.00){\oval(16.00,10.00)[]}
\put(55.00,40.00){\oval(16.00,10.00)[]}
\end{picture}
\caption{Term evaluation.}\label{TermEval}
}
\end{figure}

\subsubsection{Identification of terms}


When the standard semantics defines that during program execution
values of computed expressions are equal, we can incorporate this
information in the computation state (but it can also be left
out). {\bf Identification of terms}\/ transforms the state into a
new one incorporating this information. For example,  we know that
a value of a term representing the conditional expression of an
{\bf IF}-statement coincides with 0-ary terms representing the
constants {\bf TRUE} or {\bf FALSE} when, respectively, {\bf
THEN}-branch or {\bf ELSE}-branch is being executed. So,
identification of terms along with {\em semantic completion}
considered below provides powerful facilities to take into account
real control flow in programs.

\vskip\baselineskip
\noindent{\bf Identification of terms} \SOP{L}{t_1\equiv t_2}.

\begin{enumerate}
\item If $t_1=t_2\in L$ then $\SOP{L}{t_1\equiv t_2}=L$.

\item Let $A_{1}\DERIV{+}{G}t_1$\/ and $A_2\DERIV{+}{G}t_2$.
      We replace the nonterminal $A_2$\/ by the nonterminal
      $A_1$\/ in all rules of ${\EuScript P}$.
      If rules with an identical right side
      $B_1\rightarrow w,\ldots,B_k\rightarrow w$ have appeared, then
      a certain nonterminal from the left sides of the rules (for example $B_1$)
      must be taken and all nonterminals $B_2,\ldots,B_k$\/
      in the grammar must be replaced by it.

\item Repeat step 2 until stabilization.
      If after that we have a state $L^\prime$\/ containing inconsistent
      term equalities\footnote{There exists a wide spectrum of
      inconsistency conditions. The simplest of them is an
      equality of two different constants (see {\bf Section
      \ref{Semantic-completion}}
      for the further discussion).} then the result is \ATOP\/
      else it is a reduction of~$L^\prime$.
\end{enumerate}

\begin{figure}
\centering
{
\unitlength=0.85mm
\linethickness{0.4pt}
\begin{picture}(140.00,70.00)
\put(19.00,59.00){\framebox(6.0,6.0)[cc]{$g$}}
\put(29.00,59.00){\framebox(6.0,6.0)[cc]{$a$}}
\put(27.00,62.00){\oval(24.00,12.00)[]}
\put(9.00,40.00){\framebox(6.0,6.0)[cc]{$f$}}
\put(12.00,43.00){\circle{12.00}}
\put(27.00,40.00){\framebox(6.0,6.0)[cc]{$g$}}
\put(37.00,40.00){\framebox(6.0,6.0)[cc]{$f$}}
\put(35.00,43.00){\oval(24.00,12.00)[]}
\put(21.00,59.00){\vector(-3,-4){7.33}}
\put(23.00,59.00){\vector(1,-1){10.00}}
\put(27.00,43.00){\vector(-1,0){8.50}}
\put(9.00,21.00){\framebox(6.0,6.0)[cc]{$x$}}
\put(12.00,24.00){\circle{12.00}}
\put(32.00,21.00){\framebox(6.0,6.0)[cc]{$b$}}
\put(35.00,24.00){\circle{12.00}}
\put(12.00,40.00){\vector(0,-1){10.00}}
\put(30.00,40.00){\vector(1,-3){3.33}}
\put(40.00,40.00){\vector(-1,-3){3.33}}
\put(89.00,59.00){\framebox(6.0,6.0)[cc]{$g$}}
\put(99.00,59.00){\framebox(6.0,6.0)[cc]{$a$}}
\put(97.00,62.00){\oval(24.00,12.00)[]}
\put(89.00,40.00){\framebox(6.0,6.0)[cc]{$g$}}
\put(99.00,40.00){\framebox(6.0,6.0)[cc]{$f$}}
\put(97.00,43.00){\oval(24.00,12.00)[]}
\put(91.00,59.00){\vector(0,-1){10.00}}
\put(93.00,59.00){\vector(3,-4){7.33}}
\put(89.00,21.00){\framebox(6.0,6.0)[cc]{$x$}}
\put(99.00,21.00){\framebox(6.0,6.0)[cc]{$b$}}
\put(97.00,24.00){\oval(24.00,12.00)[]}
\put(93.00,40.00){\vector(0,-1){10.00}}
\put(102.00,40.00){\vector(0,-1){10.00}}
\put(23.00,11.00){\makebox(0,0)[cc]{$L$}}
\put(97.00,11.00){\makebox(0,0)[cc]{\SOP{L}{x\equiv b}}}
\put(84.00,37.00){\oval(14.00,8.00)[b]}
\put(79.50,39.50){\oval(5.00,7.00)[lt]}
\put(77.00,40.00){\line(0,-1){4.00}}
\put(91.00,40.00){\line(0,-1){4.00}}
\put(79.00,43.00){\vector(1,0){6.00}}
\end{picture}
\caption{Identification of values of terms.}\label{Identification}
}
\end{figure}
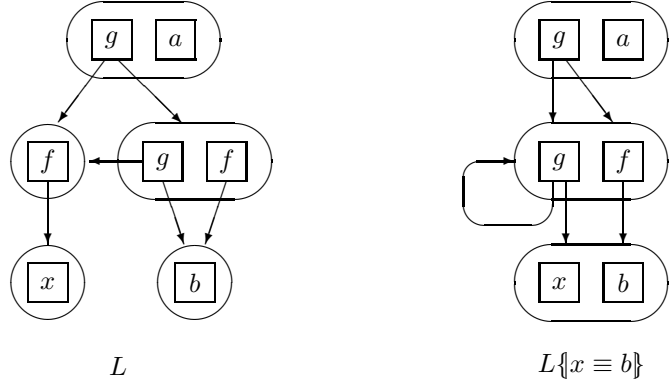

\noindent An example of identification is given in {\bf Figure \ref{Identification}}.

\begin{lem}
Identification of values of terms is a correct transformation and
the resulting state is unique.
%
%
\end{lem}

\begin{pf}

Let $\pi=\{~L_i~\vert~L_i\in\PSEQS~\}$\/ be a concrete semantic
property which holds before identification of terms $t_1$\/ and
$t_2$. If the values of $t_1$\/ and $t_2$\/ are equal in the
concrete semantics $\forall L_i ~:~ t_1=t_2\in L_i$, then they are
equal in the abstract semantics $t_1=t_2\in\alpha(\pi)$, too. If
their values are not equal, then identification gives us an
inconsistent computation state which obviously includes
\SOP{L}{t_1\equiv t_2}\/ for all $t_1$\/ and $t_2$. So, this
transformation is correct.

Identification is done in finite steps because the size of
grammar decreases at each step. Uniqueness of the resulting state
is explained by the following observation. If we have two pairs
of terms which are candidates for identification, then
identification of one of them does not close a possibility of it
for another, because we remove a duplication of the functional
symbols only. In fact, after identification of a pair of terms we
obtain a new state, including the source one, and thus other
existing identification possibilities remain. So, the order of
``merging'' of term pairs is not important for the resulting
state.
\end{pf}

\subsubsection{Semantic completion}\label{Semantic-completion}

We have not yet considered any interpretation of constants and
functional symbols. We could continue developing \ERA\/ in the
same way. As a result we shall obtain a noninterpretational
version of the analysis likewise analysis algorithms in the
program scheme theory. However, it is natural to use semantics of
primitive operations of the programming language of interest in
order to achieve better accuracy.

\ERA\/ provides us with possibilities of taking into account
properties of language constructs, and, what is especially
important, we can easily handle complexity of these manipulations.
In fact, inclusion of these properties corresponds to carrying out
some finite part of completion of the computation states by
consistent equalities. This manipulation is called {\bf semantic
completion} (about ``conjunctive/disjunctive completion'' see
\cite{CC/92a}).

\begin{figure}
\centering
{
\unitlength=0.85mm
\linethickness{0.4pt}
\begin{picture}(131.33,60.00)
\put(12.50,45.00){\oval(25.00,10.00)[]}
\put(4.00,42.00){\framebox(6.00,6.00)[cc]{$x$}}
\put(15.00,42.00){\framebox(6.00,6.00)[cc]{$+$}}
\put(52.50,45.00){\oval(25.00,10.00)[]}
\put(44.00,42.00){\framebox(6.00,6.00)[cc]{$y$}}
\put(55.00,42.00){\framebox(6.00,6.00)[cc]{{\large$\ast$}}}
\put(33.33,37.67){\oval(27.33,5.33)[b]}
\put(47.00,37.67){\vector(0,1){2.50}}
\put(19.67,42.00){\line(0,-1){4.67}}
\put(13.67,15.00){\framebox(6.00,6.00)[cc]{$z$}}
\put(16.33,42.00){\vector(0,-1){18.50}}
\put(16.67,18.17){\oval(17.33,10.33)[]}
\put(57.00,42.00){\vector(-4,-3){31.50}}
\put(53.33,15.00){\framebox(6.00,6.00)[cc]{$0$}}
\put(56.33,18.17){\oval(17.33,10.33)[]}
\put(58.67,42.00){\vector(0,-1){18.50}}
\put(34.33,5.00){\makebox(0,0)[cc]{$L$}}
\put(119.67,45.00){\oval(34.00,10.00)[]}
\put(106.67,42.00){\framebox(6.00,6.00)[cc]{$+$}}
\put(126.67,42.00){\framebox(6.00,6.00)[cc]{$z$}}
\put(116.67,42.00){\framebox(6.00,6.00)[cc]{$x$}}
\put(119.78,18.00){\oval(34.00,10.00)[]}
\put(106.78,15.00){\framebox(6.00,6.00)[cc]{$y$}}
\put(126.78,15.00){\framebox(6.00,6.00)[cc]{{\large$\ast$}}}
\put(116.78,15.00){\framebox(6.00,6.00)[cc]{$0$}}
\put(103.83,53.00){\oval(9.67,8.33)[t]}
\put(103.83,37.50){\oval(9.67,8.33)[b]}
\put(99.00,37.50){\line(0,1){15.67}}
\put(108.65,42.00){\line(0,-1){5.00}}
\put(108.67,54.0){\vector(0,-1){4.00}}
\put(111.00,42.00){\vector(1,-3){6.3}}
\put(136.00,11.50){\oval(10.67,9.00)[b]}
\put(138.50,42.00){\oval(5.67,6.67)[rt]}
\put(141.33,42.00){\line(0,-1){31.67}}
\put(138.5,45.40){\vector(-1,0){1.67}}
\put(130.67,15.00){\line(0,-1){4.33}}
\put(124.17,10.33){\oval(9.67,6.67)[b]}
\put(129.00,15.00){\line(0,-1){5.33}}
\put(119.33,9.67){\vector(0,1){3.40}}
\put(119.33,0.00){\makebox(0,0)[cc]{{\bf Co}($L$)}}
\end{picture}
\caption{Semantic completion.}\label{C-Cosure}
}
\end{figure}
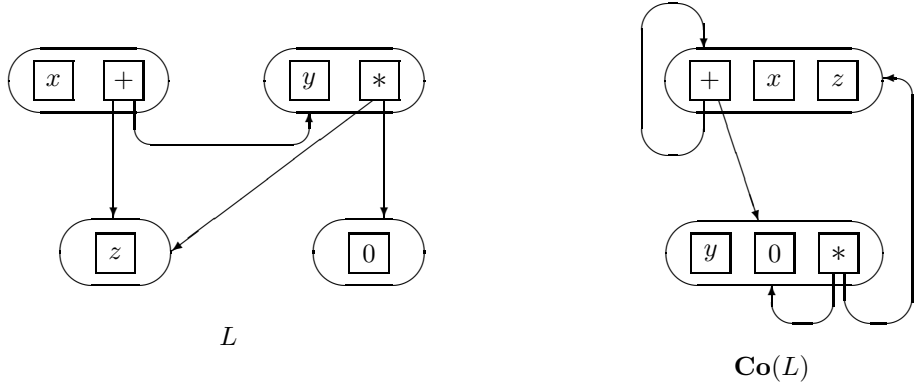

As a basic version of {\bf semantic completion} {\bf Co} we take
computations over constant and equal arguments.
%
%
When we detect that some term $t$\/ has some specific value $v$\/
then {\bf Co}($L$)$=$\SOP{L}{t\equiv v}. Also, it is possible to
apply identification involving dependencies among result of an
operation and its arguments: if $(t_1$ AND $t_2)=TRU\!E\in L$\/
then \linebreak {\bf Co}($L$)$=$\SOP{\SOP{L}{t_1\equiv
TRU\!E}}{t_2\equiv TRU\!E} etc. This identification process is
iterative because new possibilities for identification can appear
at the next steps. It is conceivable that in doing so we shall
detect inconsistency of a computation state. In this case result
of semantic completion is \ATOP.

This version can be extended by some intelligent theorem prover
inferring new reasonable equalities and checking inconsistency of
computation states. Such combining of analysis with proofs offers
powerful facilities to the analyzer (see
\cite{HeintzeJafVoi00,CousotCousot-CAV02}) and, as it was
mentioned in the previous works, this prover is reusable for
consequent (semi-)automatic processing of results of the analysis.
``Size'' of used completion can be tuned by options of
interpretability of the analyzer.

Some arithmetical errors (such as division by zero, out of type
range etc.) will appear during the semantic completion. In this
case the analyzer tells us about the error and sets the current
computation state to \ATOP. Notice that for the languages where
incomplete Boolean evaluation is admissible semantic completion
over Boolean expressions should be carefully designed especially
in presence of pointers.


An example of semantic completion is presented in {\bf Figure
\ref{C-Cosure}}. Turning back to the identification example at
{\bf Figure \ref{Identification}}, we consider the following
interpretation of constants and functional symbols: $g$\/ is the
exclusive disjunction, $f$\/ is the negation, $a$\/ is the
constant {\bf TRUE} and $b$\/ is the constant {\bf FALSE}. It is
easy to see that in this case application of semantic completion
gives us $\SOP{L}{x\equiv b}=\ATOP$.

In our analyzer we implemented an interpretational version of
\ERA\/ which uses semantic completion \Co{$L$}. Under this
approach, the definitions of the basic transformations mentioned
above are chan\-ged to the following:
\[
\begin{array}{l}
\SOP{L}{t}^\prime=\Co{\SOP{L}{t}},\\
\SOP{L}{t_1\equiv t_2}^\prime=\Co{\SOP{L}{t_1\equiv t_2}}.
\end{array}
\]
In short, we shall omit this ``interpretability'' prime.

\section{Semantic transformers}\label{STransformers}

In this section we describe semantic transformers
$\SEM{}{\centerdot}:\PSEQS\rightarrow\PSEQS$\/ corresponding to
common statements existing in imperative programming languages. If
for a statement {\bf S} an input computation state is $L$\/ then
\SEM{L}{S} is its output computation state. For all
\SEM{}{\centerdot}~ the notation
\SEM{\SEM{L}{\centerdot}}{\centerdot} means
\SEM{(\SEM{L}{\centerdot})}{\centerdot}.

\subsection{Assignment statement}

Among all program terms considered in \ERA\/ we can pick out {\em
access program terms} including array {\em elm}, record {\em fld},
and pointer {\em val} referencing and playing an important role in
determination of effect of an assignment statement. As in
\cite{Deu/94,Venet/99} our abstraction of program memory
manipulations is storeless and based on notion of memory access
paths represented by access program terms. For example, for an
address expression {\bf bar[i][j]\^{}.foo} the access term is
$\mbox{\em fld}(\mbox{\em val}(\mbox{\em elm}(\mbox{\em
elm}(bar,i),j)),foo)$.

We shall assume that no operations other than memory addressing
(for example comparisons) are allowed for structured variables
such as arrays and records. So, for the previous example neither
$a$\/ and $\mbox{\em elm}(a,i)$ ({\bf a[i]}) nor \linebreak
$\mbox{\em val}(\mbox{\em elm}(\mbox{\em elm}(a,i),j))$ ({\bf
a[i][j]\^{}}) can appear as arguments for operations other than
{\em elm} and {\em fld} respectively. This limitation allows us to
simplify definition of our assignment statement
abstraction%
\footnote{Otherwise we have either to accept that each assignment
to some structured variables destroys all equalities involving
other components of it or to implement some strategy (for example
copying) preserving  useful and safe access terms.}.

To preserve safety of the analysis we have to take into account
memory aliasing appearing in programs. Two access terms are alias
if they address the same memory location. In the general case
\ERA\/ is inadequate itself to handle precisely all kinds of
aliasing  and we should use other analyses. Next it is assumed
that for each access term $t_a$\/ we know a set \ACC{t_a} of
access terms covering a set of aliases for $t_a$\/ (may--alias
information about $t_a$). Let
\[\overline{\ACC{t_a,L({\bf G})}}=
(\ACC{t_a}\cap\CV)\cup\{f\in\FP~\vert~ A\DERIV{+}{G}f(\dots) ~\wedge~ f(\dots)\!\in\!\ACC{t_a}\}
\]
(``roots'' of memory access terms in \ACC{t_a}). Notice that
flow--insensitive approximations of alias information may cause
conservative results of \ERA. Therefore \ERA\/ and alias analyses
used in its implementation should have the same sensitivity to the
control flow.

\vskip\baselineskip
\noindent{\bf Assignment statement} \SEM{L}{v:=exp}.

\begin{enumerate}
\item In the state $L$\/ evaluate $exp$ using evaluation transformer formally defined
      earlier in {\bf Section \ref{Term-evaluation}}. Let $L^\prime$\/ be a result of the evaluation
      and let $E$\/ be a nonterminal such that $E\DERIV{}{}~exp$.
\item Perform \SOP{L^\prime}{\downarrow \overline{\ACC{v,L^\prime}}}. Do not remove $E$.
\item Add the term $v$\/ so that $E\DERIV{}{}~v$.
\end{enumerate}

Unfortunately, in some cases this abstraction of the assignment
statement fails as before. For example, this assignment
transformer corresponding to {\bf x:=x+1} and being applied to the
state $L=\{(x>0)=TRU\!E\}$\/ gives only the trivial identity
$\SEM{L}{x:=x+1}=\{x=x\}$. To improve accuracy of the analysis in
these cases we can consider ``artificial'' variables associated
with scalar variables of the program which will store previous
values of the original ones. Under this approach between first and
second steps of the assignment statement effect definition we
should insert the step
\begin{description}
\item[\dots~] Let $A\rightarrow v$\/ and $B\rightarrow v^\prime$\/ where $v^\prime$\/
is associated with $v$. Remove the second rule and add $A\rightarrow v^\prime$\/
if it is needed.
\end{description}
Under this approach we shall have $\SEM{L}{x:=x+1}=\{(x^\prime>0)=TRU\!E,x=x^\prime+1\}$\/
from where we can deduce that $x>0$\/ also.

\subsection{Other transformers}

\begin{itemize}

\item {\it Program}\/.

Given a program

{\bf
\begin{tabular}{ll}
PROGRAM;    &                     \\
~VAR~~x : T;& $(*$ variables $*)$ \\
BEGIN       &                     \\
~S          & $(*$ statements $*)$\\
END.
\end{tabular}
}

\noindent we can define the following transformer corresponding to
it

\[
\SEM{\ABOT}{PROGRAM}=\SEM{\SEM{\ABOT}{x:=\omega}}{S}
\]
where $\omega$\/ represents the indefinite value. Notice that
$\omega$\/ is not a constant.

\item {\it Empty statement}\/.
\[
\SEM{L}{~} = L
\]

\item {\it Sequence of statements}\/.
\[
\SEM{L}{S_{1}; S_{2}} = \SEM{\SEM{L}{S_{1}}}{S_{2}}
\]

\item {\it Read statement}\/.
\[
\SEM{L}{ READ(x) } = \SOP{\SOP{L}{x}}{\downarrow\overline{\ACC{x}}}
\]
Notice that if for {\em read statement} as well as for other statements
some set of user's pre-- or post--assertions represented in the form of
equalities of program terms is supplied then the analyzer can take
them into consideration to check consistency and to include in the current
computation state.

\item {\it Write statement}\/.
\[
\SEM{L}{ W\!RITE(x) } = \SOP{L}{x}
\]

\item {\it Conditional statement}\/.
\[
{\bf IF~~p~~THEN~~~S_{t}~~~ELSE~~~S_{f}~~~END}.
\]
If $L^\prime=\SOP{L}{p}$\/ then
\[
\SEM{L}{IF} = \SEM{\SOP{L^\prime}{p\equiv T\!RU\!E}}{S_{t}}\ACAP
              \SEM{\SOP{L^\prime}{p\equiv F\!ALSE}}{S_{f}}.
\]

\item {\it Cycle statement}\/.

{\bf
\begin{tabular}{lll}
CYCLE  &     &                        \\
       &  S  & $(*$ body of cycle $*)$\\
END    &     &                        \\
\end{tabular}
}

where {\bf S} is a composed statement that possibly contains occurrences of
exit-of-cycle statements {\bf EXIT$_k$}. When the sequence
\[
L_{0}=L\mbox{,~~}L_{n} = \SEM{L_{n-1}}{S}\mbox{~~for~~} n > 0
\]
becomes stabilize
\[
\SEM{L}{CYCLE} = \ACAP_{k}E_{k}
\]
where $E_{k}$\/ is a stationary entry state for ${\bf EXIT_{k}}$.
If this process does not become stabilize then some widening
operator should be used (see Section \ref{WideningOp}).

\item {\it Halt, exit and return statements}\/.
\[
\SEM{L}{EXIT}~=~\SEM{L}{RETURN}~=~\ATOP
\]

\item {\it Call of function}\/. We assume that return of results
of function calls is implemented as an assignment to variables
having the same names as invoked functions (connection with their
call sites should be taken into account). The function bodies may
contain {\bf RETURN} statements as well.

{\bf
\begin{tabular}{ll}
FUNCTION F(x: T$_1$) : T;  &           \\
~~VAR~~y: T$_2$;           & $(*$ local variables $*)$ \\
BEGIN                      &                     \\
~~S                        & $(*$ statements $*)$\\
END.
\end{tabular}
}
\[
\SEM{L}{FUNCTION}=\SEM{\SEM{L}{x:=e;y:=\omega;F:=\omega}}{S}\ACAP
R
\] where $e$\/ is a factual parameter of the function (see the
note for {\bf READ} statement) and $R=\ACAP R_k$\/ is intersection
of stationary entry states for return statements in {\bf S} if
they exist and $R=\ATOP$\/ otherwise. In the term being evaluated
the result of a function call is represented by $F$.


\begin{figure}[t]
\centering {\small \unitlength=0.70mm
\linethickness{0.4pt}
\begin{picture}(100.00,50.00)
\put(10.00,30.00){\oval(22.00,12.00)[]}
\put(3.0,27.90){\framebox(4.2,4.2)[cc]{$x$}}
\put(12.80,27.90){\framebox(4.2,4.2)[cc]{$y$}}
\put(10.00,20.00){\makebox(0,0)[cc]{$L$}}
\put(75.00,40.00){\oval(30.00,12.00)[]}
\put(65.0,37.90){\framebox(4.2,4.2)[cc]{$x$}}
\put(72.90,37.90){\framebox(4.2,4.2)[cc]{\tiny $+$}}
\put(81.20,37.90){\framebox(4.2,4.2)[cc]{\tiny $F$}}
\put(74.00,37.90){\vector(-1,-1){12.00}}
\put(76.00,37.90){\vector(1,-1){12.00}}
\put(60.00,20.00){\circle{12.00}}
\put(90.00,20.00){\circle{12.00}}
\put(57.9,17.90){\framebox(4.2,4.2)[cc]{$y$}}
\put(87.90,17.90){\framebox(4.2,4.2)[cc]{$1$}}
\put(75.0,5.00){\makebox(0,0)[cc]{$\SEM{L}{x := F(x)}$}}
\end{picture}
\caption{Function call for $F(a)=a+1$.} }
\end{figure}
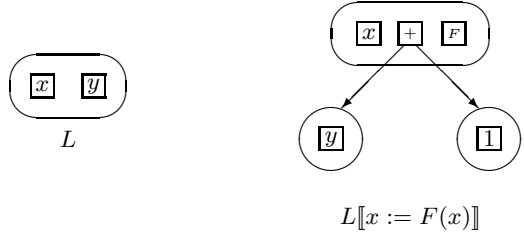

\end{itemize}


%





\section{Widening operator and convergence of the analysis}\label{WideningOp}

Our abstract semantic domain does not satisfy the {\em
(descending) chain condition} and therefore it requires some {\em
widening operator} \cite{CC/77,CC/92b,CC/92a}. To guarantee the
convergence of the abstract interpretation we should use a dual
widening operator:
\begin{itemize}
\item $\forall x,y\in\PSEQS~~\Rightarrow~~
      x~\AINCL~x\tilde{\bigtriangledown}y~~\&~~
      y~\AINCL~x\tilde{\bigtriangledown}y$,

\item for all decreasing chains $x_0\AINCL x_1\AINCL\ldots$,
      the decreasing chains defined by
      $y_0=x_0,\ldots,y_{i+1}=y_i\tilde{\bigtriangledown}x_{i+1},\dots$\/
      are not strictly decreasing.
\end{itemize}
The iteration sequence with widening is convergent and
its limit is a sound approximation of the fixpoint.

\subsection{Widening operator for \ERA--grammars}

Infinite chains can appear because corresponding languages have
common infinite subsets generated by cyclic derivations in
grammars. The source of that in \ERA\/ is {\bf term
identification}. We can avoid this problem by imposing the
constraint that grammars must be acyclic. Within the semilattice
\ASDE, the subsemilattice of finite languages\footnote{Notice that
sets of term equalities of a special form corresponding to these
languages were used by V. Sabelfeld to develop effective
algorithms of recognizing equivalence for some classes of program
schemata \cite{Sa/79,Sa/80}.} generated by such grammars satisfies
the chain condition, but such languages are not expressive enough.
Our solution is the following. Grammars are not originally
restricted but if in course of abstract interpretation the grammar
size becomes greater than some parameter, then ``harmful'' cycles
must be destroyed. To this end we remove grammar rules
participating in cyclic derivations. Correctness of this
approximation of intersection follows from {\bf Lemma
\ref{DeleteTerms}}.


Detecting such rules is no simpler than the
``mi\-ni\-mum-\-feed\-back-arc/ver\-tex-set'' problem ({\bf MFAS}
or {\bf MFVS}) if we consider the grammars as directed graphs.
These sets are the smallest sets of arcs or vertices,
respectively, whose removal makes a graph acyclic. We suppose that
the ``feedback vertices'' choice is more natural for our purposes.
In the general case this problem is ${\EuScript NP}$--hard, but
there are approximate algorithms that solve this problem in
polynomial \cite{Speckenmeyer/89} or even linear \cite{Rosen/82}
time. Consideration of weighted digraphs makes it possible to
distinguish grammar rules with respect to their worth for accuracy
of the analysis algorithm. However, perspectives of this approach
are not clear now for complexity/precision reasons of such
algorithms. For example, \cite{ENSS/95} proposes an algorithm for
weighted {\bf FVS}-problem requiring $O(n^2\mu(n)\log^2n)$\/ time
where $\mu(n)$\/ is complexity of matrix multiplication.

\begin{figure}[h]
\centering
\unitlength=0.825mm
\linethickness{0.4pt}
\begin{picture}(160.00,80.00)
\put(12.50,45.00){\oval(25.00,10.00)[]}
\put(5.00,42.00){\framebox(6.00,6.00)[cc]{$g$}}
\put(15.00,42.00){\framebox(6.00,6.00)[cc]{$r$}}
\put(44.00,42.00){\framebox(6.00,6.00)[cc]{$h$}}
\put(54.00,42.00){\framebox(6.00,6.00)[cc]{$y$}}
\put(64.00,42.00){\framebox(6.00,6.00)[cc]{$f$}}
\put(57.00,45.00){\oval(34.00,10.00)[]}
\put(41.80,38.33){\oval(47.60,6.67)[b]}
\put(65.60,42.00){\line(0,-1){4.67}}
\put(66.67,37.33){\line(0,1){0.00}}
\put(18.00,39.75){\vector(0,1){0.20}}
\put(18.00,37.33){\line(0,1){2.00}}
\put(39.50,45.00){\vector(1,0){0.20}}
\put(21.00,45.00){\line(1,0){18.33}}
\put(29.50,15.00){\oval(25.00,10.00)[]}
\put(22.00,12.00){\framebox(6.00,6.00)[cc]{$h$}}
\put(32.00,12.00){\framebox(6.00,6.00)[cc]{$z$}}
\put(28.50,20.33){\vector(3,-4){0.20}}
\multiput(10.00,42.00)(0.12,-0.14){150}{\line(0,-1){0.14}}
\put(21.00,20.33){\vector(2,-3){0.20}}
\multiput(6.00,42.00)(0.12,-0.17){126}{\line(0,-1){0.17}}
\put(35.33,20.33){\vector(-1,-2){0.20}}
\multiput(47.00,42.00)(-0.12,-0.22){98}{\line(0,-1){0.22}}
\put(14.49,11.80){\oval(21.00,13.00)[b]}
\put(4.00,39.75){\vector(0,1){0.20}}
\put(4.00,11.90){\line(0,1){27.67}}
\put(24.00,72.00){\framebox(6.00,6.00)[cc]{$x$}}
\put(34.00,72.00){\framebox(6.00,6.00)[cc]{$f$}}
\put(44.00,72.00){\framebox(6.00,6.00)[cc]{$a$}}
\put(37.00,75.00){\oval(34.00,10.00)[]}
\put(75.00,38.33){\oval(12.00,6.67)[b]}
\put(69.00,42.00){\line(0,-1){4.00}}
\put(78.00,72.67){\oval(6.00,5.33)[rt]}
\put(81.00,37.33){\line(0,1){35.67}}
\put(78.00,75.33){\vector(-1,0){23.67}}
\put(35.00,72.00){\vector(-1,-1){22.00}}
\put(38.33,72.00){\vector(3,-4){16.25}}
\put(40.00,0.00){\makebox(0,0)[cc]{{\small$L({\bf G})$}}}
\put(117.50,45.00){\oval(25.00,10.00)[]}
\put(110.00,42.00){\framebox(6.00,6.00)[cc]{$g$}}
\put(120.00,42.00){\framebox(6.00,6.00)[cc]{$r$}}
\put(149.00,42.00){\framebox(6.00,6.00)[cc]{$y$}}
\put(159.00,42.00){\framebox(6.00,6.00)[cc]{$h$}}
\put(171.67,37.33){\line(0,1){0.00}}
\put(144.50,45.00){\vector(1,0){0.20}}
\put(126.00,45.00){\line(1,0){18.33}}
\put(136.50,15.00){\oval(15.00,10.00)[]}
\put(133.67,12.00){\framebox(6.00,6.00)[cc]{$z$}}
\put(123.67,72.00){\framebox(6.00,6.00)[cc]{$x$}}
\put(133.67,72.00){\framebox(6.00,6.00)[cc]{$f$}}
\put(143.67,72.00){\framebox(6.00,6.00)[cc]{$a$}}
\put(136.67,75.00){\oval(34.00,10.00)[]}
\put(157.50,45.00){\oval(25.00,10.00)[]}
\put(135.00,72.00){\vector(-3,-4){16.25}}
\put(138.00,72.00){\vector(3,-4){16.25}}
\put(111.67,42.00){\vector(3,-4){18.00}}
\put(161.00,42.00){\vector(-3,-4){18.00}}
\put(114.67,42.00){\vector(1,-1){22.00}}
\put(136.67,0.00){\makebox(0,0)[cc]{{\small$L({\bf G})\setminus_{\bf fvs}$}}}
\end{picture}
%
%
\epsfig{file=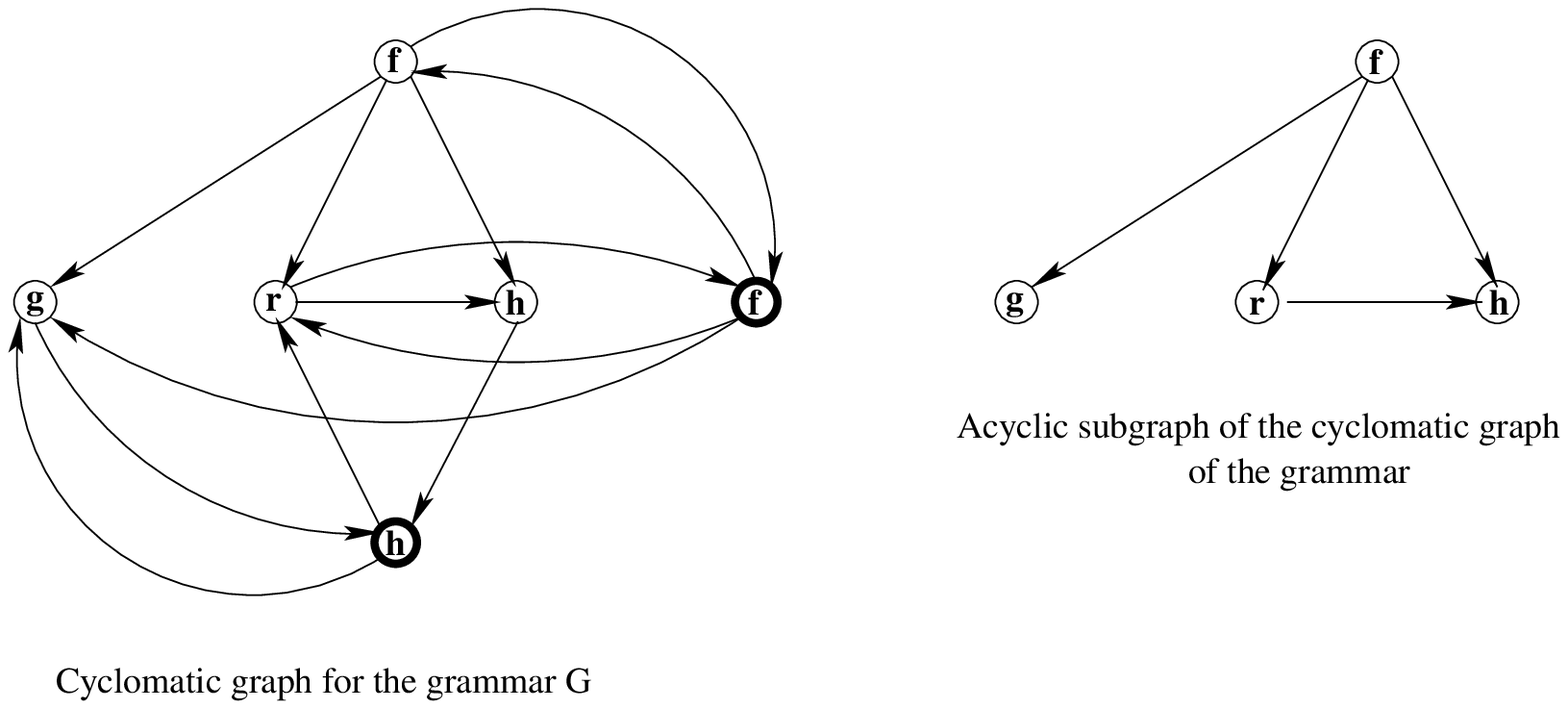,width=390pt}
\caption{{\bf FVS}--transformation.}\label{MFVS}
\end{figure}

Our widening operator for the analysis of equality relationships
is defined in the following way. A vertex set of the cyclomatic
graph\footnote{A graph represents cyclic derivations in the
grammar. The author did not find another appropriate name for this
object in the Computer Science and Discrete Mathematics
literature.} of the grammar {\bf G} is a set of functional symbols
existing in {\bf G}. An arc $(f,g)$\/ belongs to its arc set if
{\bf G} contains rules $A\rightarrow f(\ldots,B,\ldots)$\/ and
$B\rightarrow g(\ldots)$. A transformation of a cyclomatic graph
which involves detecting some {\bf FVS} (it can be an upper
approximation of a minimal feedback set) and removing all vertices
from the {\bf FVS} is said to be an {\bf FVS}--transformation (an
example is shown in {\bf Figure \ref{MFVS}}). Let \MFVS{L}\/ be a
language obtained from $L$\/ by {\bf FVS}--transformation applied
to the grammar generating $L$. We define
\[
L(\mbox{\bf G}_1)\DWID L(\mbox{\bf G}_2) =
\left\{\begin{array}{ll}
  \MFVS{L(\mbox{\bf G}_1)}~\ACAP~L(\mbox{\bf G}_2) &
  \mbox{~if~~}\vert\mbox{\bf G}_2\vert\geq\vert\mbox{\bf G}_1\vert>d,\\
  L(\mbox{\bf G}_1)~\ACAP~\MFVS{L(\mbox{\bf G}_2)} &
  \mbox{~if~~}\vert\mbox{\bf G}_1\vert>\vert\mbox{\bf G}_2\vert>d,\\
  L(\mbox{\bf G}_1)~\ACAP~L(\mbox{\bf G}_2)        & \mbox{~otherwise,}
\end{array}\right.
\]
where $d$\/ is a user-defined parameter. It is reasonable to
choose this parameter, depending on number of variables of
analyzed programs, as a linear function with a small factor of
proportionality. Notice that in this case the lengths of appearing
chains linearly depend on number of variables living
simultaneously.

\subsection{Divergence of the analysis}

Is the widening operator, being rather complex, really needed for
the analysis of equality relationships? Are there programs which,
being analyzed, generate infinite chains of semantic properties?
It should be mentioned that constructing such program examples has
been a problem for a long time. In \cite{Em/96} we stated our
belief that their existence seems hardly probable. These attempts
failed, because they concentrated on constructing an example with
completely non-interpretable functional symbols, i.e. in the frame
of the ``pure'' theory of program schemata.

As already noticed, we can widely vary the interpretability of the
analysis algorithm. In order to construct an example, it will
suffice to use the following rule of completion:
\[
\mbox{if } (t_1=t_2)=TRU\!E\in L \mbox{ then {\bf Co}}(L)=\SOP{L}{t_1\equiv t_2}.
\]

We consider the following example:

\vskip\baselineskip
\begin{center}
\hspace*{-7.5mm}\begin{tabular}{ccc}
\begin{tabular}{l}
     ...\\\vspace*{-2pt}
     x:=f(y);\\\vspace*{-2pt}
     \KW{if}~ f(x)=f(y) ~\KW{then}\\\vspace*{-2pt}
~~       \KW{while}~ y=f(g(y)) ~\KW{do}\\\vspace*{-2pt} %
~~~~     y:=g(y)\\\vspace*{-2pt} ~~%
       \KW{end}\\\vspace*{-2pt}
     \KW{end}\\\vspace*{-2pt}
     ...\\
\hspace*{6mm}{\bf program scheme}
\end{tabular}
&\hspace*{7.5mm}&
\begin{tabular}{l}

     ...\\\vspace*{-2pt}
     x:=sign(y);\\\vspace*{-2pt}
     \KW{if}~ sign(x)=sign(y) ~\KW{then}\\\vspace*{-2pt}
~~       \KW{while}~ y=sign(abs(y)) ~\KW{do}\\\vspace*{-2pt} %
~~~~     y:=abs(y)\\\vspace*{-2pt} ~~%
         \KW{end}\\\vspace*{-2pt}
     \KW{end}\\\vspace*{-2pt}
    ...\\
\hspace*{5mm}{\bf ``real-world'' program}
\end{tabular}\\
\end{tabular}
\end{center}
\vskip\baselineskip

\begin{figure}[t]
\centering
{
\unitlength=0.85mm
\linethickness{0.4pt}
\begin{picture}(140,100.00)
\put(4.00,81.00){\framebox(4.00,4.00)[cc]{$x$}}
\put(10.00,81.00){\framebox(4.00,4.00)[cc]{$f$}}
\put(12.00,81.00){\vector(0,-1){13.00}}
\put(16.00,81.00){\framebox(4.00,4.00)[cc]{$f$}}
\put(12.00,83.00){\oval(22.00,10.00)[]}
\put(22.50,76.00){\oval(9.00,10.00)[b]}
\put(22.50,90.00){\oval(9.00,9.00)[t]}
\put(18.00,90.00){\vector(0,-1){2.00}}
\put(27.00,90.00){\line(0,-1){14.00}}
\put(18.00,81.00){\line(0,-1){5.00}}
\put(10.00,61.00){\framebox(4.00,4.00)[cc]{$y$}}
\put(12.00,63.00){\circle{10.00}}
\put(12.00,50.00){\makebox(0,0)[cc]{$L_e$}}

\put(54.00,81.00){\framebox(4.00,4.00)[cc]{$x$}}
\put(60.00,81.00){\framebox(4.00,4.00)[cc]{$f$}}
\put(62.00,81.00){\vector(0,-1){13.00}}
\put(66.00,81.00){\framebox(4.00,4.00)[cc]{$f$}}
\put(62.00,83.00){\oval(22.00,10.00)[]}
\put(72.50,76.00){\oval(9.00,10.00)[b]}
\put(72.50,90.00){\oval(9.00,9.00)[t]}
\put(68.00,90.00){\vector(0,-1){2.00}}
\put(77.00,90.00){\line(0,-1){14.00}}
\put(68.00,81.00){\line(0,-1){5.00}}
\put(62.00,63.00){\oval(18.00,10.00)[]}
\put(56.00,61.00){\framebox(4.00,4.00)[cc]{$y$}}
\put(63.00,61.00){\framebox(4.00,4.00)[cc]{$f$}}
\put(65.00,61.00){\vector(0,-1){13.00}}
\put(65.00,43.00){\circle{10.00}}
\put(63.00,41.00){\framebox(4.00,4.00)[cc]{$g$}}
\put(65.00,41.00){\line(0,-1){5.00}}
\put(69.50,36.00){\oval(9.00,9.00)[b]}
\put(74.00,36.00){\line(0,1){24.00}}
\put(71.00,60.00){\oval(6.00,6.00)[rt]}
\put(71.50,63.00){\vector(-1,0){0.75}}
\put(62.00,29.00){\makebox(0,0)[cc]{$L_1$}}

\put(114.00,83.00){\oval(22.00,10.00)[]}
\put(106.00,81.00){\framebox(4.00,4.00)[cc]{$x$}}
\put(112.00,81.00){\framebox(4.00,4.00)[cc]{$f$}}
\put(114.00,81.00){\vector(0,-1){13.00}}
\put(118.00,81.00){\framebox(4.00,4.00)[cc]{$f$}}
\put(124.50,76.00){\oval(9.00,10.00)[b]}
\put(124.50,90.00){\oval(9.00,9.00)[t]}
\put(120.00,90.00){\vector(0,-1){2.00}}
\put(129.00,90.00){\line(0,-1){14.00}}
\put(120.00,81.00){\line(0,-1){5.00}}
\put(103.00,56.00){\dashbox(22.00,14.00){}}
\put(112.00,61.00){\framebox(4.00,4.00)[cc]{$f$}}
\put(114.00,63.00){\circle{10.00}}
\put(114.00,61.00){\vector(0,-1){13.00}}
\put(114.00,43.00){\oval(22.00,10.00)[]}
\put(106.00,41,00){\framebox(4.00,4.00)[cc]{{\footnotesize$\!g^*$}}}
\put(108.00,41.00){\line(0,-1){5.00}}
\put(103.50,37.00){\oval(9.00,9.00)[b]}
\put(99.00,37.00){\line(0,1){24.00}}
\put(102.00,61.00){\oval(6.00,6.00)[lt]}
\put(102.00,64.00){\vector(1,0){7.0}}
\put(112.00,41.00){\framebox(4.00,4.00)[cc]{$f$}}
\put(118.00,41.00){\framebox(4.00,4.00)[cc]{$y$}}
\put(114.00,41.00){\vector(0,-1){13.00}}
\put(114.00,23.00){\circle{10.00}}
\put(112.00,21.00){\framebox(4.00,4.00)[cc]{$g$}}
\put(114.00,21.00){\line(0,-1){5.00}}
\put(121.25,16.00){\oval(14.50,9.00)[b]}
\put(128.50,16.00){\line(0,1){24.00}}
\put(125.50,40.00){\oval(6.00,6.00)[rt]}
\put(125.50,43.00){\vector(-1,0){0.75}}
\put(114.00,3.00){\makebox(0,0)[cc]{$L_2$}}
\end{picture}
\caption{The case of divergence of the analysis.}\label{RealChain} }
\end{figure}
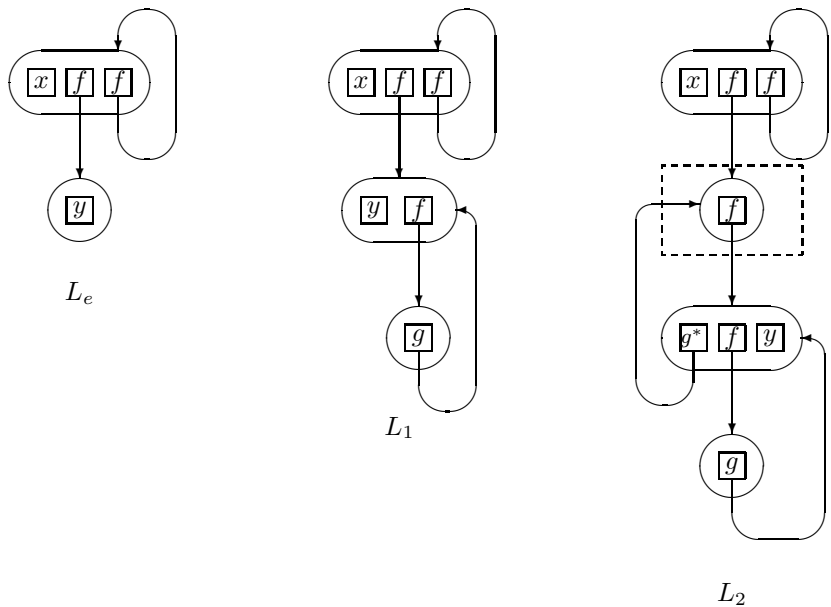

The properties computed at the body entry belong to an infinite
decreasing chain. On {\bf Figure \ref{RealChain}} a state $L_e$\/
describes properties valid before cycle execution; states $L_1$\/
and $L_2$\/ describe properties at the entry of the cycle body for
first and second iteration, respectively. It is easy to see that
$L_1\ACAP L_2$\/ coincides with $L_2$\/ except in the equality
relationships containing terms generated by $g^*$. Therefore every
time we obtain the next state, the functional element $g^*$\/ is
absent and the sub-net placed in the dashed box will repeat more
and more times.

To obtain a ``real-world'' program from this program scheme, we
interpret the functional symbols in the following manner: $f\equiv
sign$\/ and $g\equiv abs$. We would like to point out the
following interesting property. Execution of this piece of code
(i.e., its behavior determined with the standard semantics)
diverges only for two values of \OPR{y}: 0 and 1. At the same time
the analysis algorithm (i.e. execution of the piece of code under
our nonstandard semantics) is always divergent on condition that a
widening operator is not used and the assumption on interpretation
mentioned above holds.

Is this program actually real-world? The reader should decide that
by himself but we would like to underline the
following\footnote{The penetrating reader may notice that this
program to be considered as human written is really stupid (it can
be slightly intellectualized if we interpret $g$\/ as ``{\em add
1}''). But for automatic generators of programs such a code does
not seem improbable.}. On the one hand, the interpretability of
the analysis algorithm can be varied in wide ranges and, on the
other, we are not able to prove formally the impossibility of such
 behavior of the analyzer under the considered interpretation.
So, we can choose either a lean analysis using acyclic grammars
only or another one using arbitrary grammars and some widening
operator.

\subsection{Complexity of the analysis}\label{ComplexityERA}

In \cite{Em/96} we pointed out the following upper bound on time
for the algorithm of \ERA: $O(nmG^2_{max})$, where $n$\/ is
program size, $m$\/ is maximum of number of program variables
existing at a time, and $G_{max}$\/ is maximum of sizes of
grammars appearing in course of the analysis. Due to our
construction of widening operator (namely, choice of the parameter
$d$\/ linearly depended on $m$) we can assume that $G_{max}=O(m)$.

This bound can be deduced with help of {\bf Theorem 6} in
\cite{Bou/93a}. The theorem states that for the recursive strategy
of chaotic iteration maximum complexity is
\[
h\cdot\sum_{c\in C}\delta(c) \hspace*{10mm}
\left(\strut\leq h\cdot\vert C\vert\cdot\vert W\vert\right),
\lefteqn{\hspace*{2.5cm}{\bf(**)}}
\]
where $h$\/ is maximum length of increasing chains built by a
widening operator, $C$\/ is set of control program points,
$\delta(c)$\/ is depth of the control point $c$\/ in hierarchy of
nested strongly connected components of the control flow graph
containing $c$, and $W$\/ is set of vertices  where a widening
operator is applied during the analysis.

For well-structured programs we can assume that maximum depth of
nested loops does not depend on program size and is bounded by
some constant. By ${\bf(**)}$\/ we conclude that number of
algorithm steps does not exceed $O(nm)$. Since time complexity of
all operations used in the analysis are estimated by $G^2_{max}$\/
we obtain $O(nmG^2_{max})$\/ (or $O(nm^3)$) upper bound. Notice
that to improve the results of the analysis it is possible to use
rich semantic completion and more precise {\bf FVS}--algorithms
that have more than quadratic time complexity.

However, experimental results show that an approximation of a
fixed point for the heads of cycle bodies is usually attained
after at most two iterations and time complexity of the analysis
is proportional to $nG_{max}$. Also, the user can turn off
checking a threshold after which widening is started. In this
case, he (consciously) admits some chance that the analysis
diverges but we believe that this chance is not too big.

It is easy to see that the space complexity of the equality relationship
analysis is $O(nG_{max})$\/ and it is  essentially depended
on the number of variables.
We estimate the actual space requirements as 1.5--2.0 Mb
per 1000 program lines for middle-size programs.

\section{Processing of invariants and experimental results}\label{Experiments}

\subsection{Usage of \ERA--invariants}

\ERA\/ produces some set of invariants involving program terms
which can be useful at different steps of program development and
processing: debugging, verification (for that the invariants are
interesting themselves), specialization, and optimization. The
automatic prover mentioned above can be used at step of
post-processing results of the analysis.

We notice that the analyzer can tell the user useful information both at the stage of
analyzing (this means that it is possible that there exist execution traces where
such computational states appear; we mark with $+$\/ properties which can be detected
at this stage) and at the stage of processing of results of the analysis
(these properties hold for each execution trace leading to this program point).
We shall briefly list some program properties that can be extracted from computation
states $L_{in}$\/ and $L_{out}$\/ being for a statement {\bf S} the input and output
states respectively.

\begin{description}
\item[$+$] {\em Variable $x$\/ has an indefinite value}\/
           if $L_{in}$\/ contains $x=\omega$.


\item[$+$] {\em Error in evaluation of an expression}: division by zero,
           out of type ranges, nil--pointer dereferencing, etc.

\item[$-$] {\em {\bf S} is inaccessible} if $L_{in}=\ATOP$. This information
           can correspond to different properties of program execution:
           potentially infinite cycles and recursive calls, dead branches of
           conditional statements, useless procedural definitions, etc.

\item[$-$] {\em Assignment statement {\bf v:=exp} is redundant} if
           $L_{out}$\/ contains $v=v^\prime$\/ where $v^\prime$\/ is the
           variable associated with $v$.

\item[$-$] {\em Unused definitions} (constants, variables, types).

\item[$-$] {\em Constant propagation}. Notice that \ERA\/ can detect that an expression
           is constant not only when constants for all variables in this expression
           are known.

\item[$-$] More general: for some expression {\em there exists an expression that is
           equal to the original one and calculated more efficiently}\/
           (with respect to an given criterion
           time/space and target computer architecture).
\end{description}
Obviously, this list is not complete and there are many other properties
which can be extracted from the invariants. For example, we can consider
systems of equations/inequalities contained in the gathered invariants
and try to solve them to derive more precise ranges for values of expressions
or the inconsistency of this computational state.

Apart from the automatic mode when invariants are processed
automatically, we provide an interactive mode to visualize results
of our analysis in a hypertext system. {\em HyperCode} presented
in \cite{BuKo/98,BBEF/2001} is an open tunable system for
visualization of properties of syntactic structures. There are two
cases: visualization of all properties detected in the automatic
mode and the user-driven processing and visualization of
properties.

The experiments show that not all program properties of interest
can be automatically extracted out of the computed invariants. It
is not judicious to consider many particular cases and to hardly
embed them into the system. Instead, the system facilitates the
specification of the user request with some friendly interface. He
chooses a program point and an expression and obtains those and
only those equality relationships, valid at this point, where this
expression occurs as a super- or sub-term.

\subsection{Program examples}

An example of a program is presented below. The properties detected by the
analyzer are indicated in comments.

\vskip\baselineskip
\begin{tabular}{ll}
\KW{var}  x,y,z: \KW{integer}; &\\\vspace*{-2pt}
\KW{proce}\lefteqn{\mbox{\rm\KW{dure} P(a,b:
\KW{integer}):\KW{integer};}}&\\\vspace*{-2pt} \KW{begin}
&(*parameters are always equal*)\\\vspace*{-2pt} ~~  \KW{return}
a+b                           &(*expression can be simplified:
{\bf 2*a}*)\\\vspace*{-2pt} \KW{end}  P; &\\\vspace*{-2pt}
\KW{begin} &\\\vspace*{-2pt} ~~   Read(x); &\\\vspace*{-2pt} ~~
\KW{while}~  x$\leq$0  ~\KW{do} &\\\vspace*{-2pt} ~~~~   Read(x);
&\\\vspace*{-2pt} ~~~~   x := x+1; &\\\vspace*{-2pt} ~~~~   z :=
x+z; &(*variable {\bf z} might be uninitialized*)\\\vspace*{-2pt}
~~~~ y := x+1;                         &\\\vspace*{-2pt} ~~~~
\KW{if}~  x=0  ~\KW{then}                   &\\\vspace*{-2pt}
~~~~~~   z := y;                         &(*r-value can be
simplified: {\bf z:=1}*)\\\vspace*{-2pt} ~~~~   \KW{else}
&\\\vspace*{-2pt} ~~~~~~   z := x+1; &(*r-value can be simplified:
{\bf z:=y}*)\\\vspace*{-2pt} ~~~~~~ x := y; &\\\vspace*{-2pt} ~~~~
\KW{end}; &\\\vspace*{-2pt} ~~~~   Write( P(y,z) ) &(*call can be
transformed: {\bf Write(2*y)}*)\\\vspace*{-2pt} ~~ \KW{end};
&\\\vspace*{-2pt} ~~   x := z \KW{div} (y - z); &(*arithmetical
error*)\\\vspace*{-2pt} ~~   Write(x) &(*inaccessible
point*)\\\vspace*{-2pt}
\KW{end.}\\
\end{tabular}
\vskip\baselineskip

\noindent On basis of the analysis, this program can be
transformed into the following:

\vskip\baselineskip
\begin{center}
\begin{tabular}{l}
\KW{var}  x:\KW{integer};\\\vspace*{-2pt}
\KW{begin}\\\vspace*{-2pt} ~~   Read(x);\\\vspace*{-2pt} ~~
\KW{while}~  x$\leq$0  ~\KW{do}\\\vspace*{-2pt} ~~~~   Read(x);~~
x := x+2;~~   Write(2*x)\\\vspace*{-2pt} ~~
\KW{end}\\\vspace*{-2pt} ~~ \KW{ERROR\_EXCEPTION};\\\vspace*{-2pt}
\KW{end.}\\
\end{tabular}
\end{center}
\vskip\baselineskip

\begin{table}
\begin{center}
\begin{tabular}{|| l || c | c | c || c | c | c ||}
\hline\hline
\raisebox{-2ex}{program}
          & \multicolumn{3}{c||}{length (lines)} & \multicolumn{3}{c||}{size (bytes)}\\
            \cline{2-4}                           \cline{5-7}
          & {\em M2Mix} & \ERA &  improv.        & {\em M2Mix} & \ERA & improv.\\
\hline
\bf KMP       &    167      & 133 &  20.35\%        &    2996     & 2205& 26.4\%\\
\bf Lambert   &    361      & 326 &  9.7\%          &    6036     & 2564& 57.5\%\\
\bf Automaton &    37       & 35  &  5.4\%          &    969      & 926 & 4.5\%\\
\bf Int$_{Fib}$&   87       & 77  &  11.5\%         &    1647     & 1432& 13.05\%\\
\bf Ackerman  &    64       & 62  &  3.1\%          &    1384     & 1322& 4.5\%\\
\hline
          & \multicolumn{2}{r|}{average}&10.01\%&\multicolumn{2}{r|}{average}&21.19\%\\
\hline\hline
\end{tabular}
\caption{Comparison \ERA\/ and {\em
M2Mix}.}\label{M2Mix-analysis}\label{TableCMP}
\end{center}
\end{table}

In {\bf Table \ref{M2Mix-analysis}} we present some results of
optimization based on our analysis of residual programs generated
by {\em M2Mix} specializer \cite{BuKo/96,Ko/95:phd}. To compile
these examples, we used  XDS Modula/Obe\-ron compiler v.2.30
\cite{XDS/2000}. The following programs have been investigated:

\begin{itemize}
\item {\bf KMP}       --- the ``na{\"\i}ve'' matching algorithm
                      specialized with respect to some pattern;
                      the residual program is comparable to
                      Knuth, Morris, and Pratt's algorithm in efficiency.
                      (see also {\bf Appendix}).

\item {\bf Lambert}   --- a program drawing Lambert's figure
                      and specialized with respect to number of points.

\item {\bf Automaton} --- an interpreter of a deterministic finite-state
                      automaton specialized with respect to some language.

\item {\bf Int$_{Fib}$} --- an interpreter of MixLan \cite{Ko/95:phd}
                      specialized with respect to a program computing
                      Fibonacci numbers.

\item {\bf Ackerman}  --- a program computing some values of Akcerman's
                      function and specialized with respect to its
                      first argument.
\end{itemize}

Let us comment briefly on the obtained results. Reducing length of
a program can be considered as reducing number of operators and
declarations. In these examples the optimizing effect was
typically attained by the removal of redundant assignments, dead
operators, unused variables and the reduction of operator
strength. The only exception is {\bf KMP} program characterized by
high degree of polyvariance (roughly speaking it means presence of
deep--nested conditional statements) and an active usage of array
references. Here some IF--statements with constant conditions and
redundant range checks were eliminated. Notice that the last
optimizing transformation is very important for Modula--like
languages where such checks are defined by the language standard.
Such notable optimizing effect for the {\bf Lambert} program is
explained by deep reduction of power of floating-point operations
which cannot be achieved by optimizing techniques now used in
compilers. Since {\bf Automaton} and {\bf Ackerman} programs are
quite small, their optimization gives conservative results.
However, they would be better for the {\bf Ackerman} program if
the implementation of \ERA\/ were context-sensitive. Substantial
speed-up of these optimized programs was not obtained (it was less
than 2\%) and this is not surprising since the great bulk of
specializers take it as a criterion of optimality.

These experiments show that an average reduction of size of
residual programs is 20--25\%. Because the case of {\bf KMP}
program seems to be the most realistic\footnote{Unfortunately our
experiments were not exhaustive enough since the partial
evaluation is not involved yet in real technological process of
the software development and hence finding large resudial programs
is a hard problem.}, we suppose that such improvement can be
achieved in practice for real--world programs and it will be
increased for large residual programs with a
high degree of polyvariance 
and active usage of arrays and float-point arithmetics. It is the
author opinion that the analysis of automatically generated (from
high--level specifications as well) programs is the most promising
direction of its application, especially in the context-sensitive
implementation of \ERA.

\section*{Acknowledgement} The author wishes to thank M.A.
Bulyonkov, P. Cousot, R. Cousot, and V.K. Sabelfeld for support,
useful discussions and remarks.



\newpage

\section*{Appendix. Analysis of KMP}

The appendix presents results of application of \ERA\/ to {\bf
KMP} program generated by the specializer {\em M2Mix}. This
program is a specialization of a program implementing the
na{\"\i}ve pattern matching {\bf Match(p,str)} with respect to the
pattern {\bf p="ababb"}. The invariants are written as comments at
program points where they hold.
%
We can conclude that:

\begin{itemize}
\item The target string necessarily ends with "\#" and the variable
      {\bf ls} is equal to the string length (line 10).

\item Every time when some element of {\bf str}
      (lines 20, 31, 42, 53, 58, 67, 72, 78, 86, 108, 113, 129, 140, 151, 162)
      is used in second {\bf LOOP}, the value of its index
      expression does not exceed the value of the variable {\bf ls}.
      The same is true for the value of a variable before the
      increment statements {\bf INC(s)}
      (lines 23, 34, 45, 62, 81, 89, 94, 98, 103, 116, 120, 132, 143, 154, 165).
      Therefore, it suffices to check that a value of {\bf ls}
      is not beyond the ranges determined by the type {\bf \_TYPE354a04}
      during input of the target string (line 8). So, in the second cycle
      all range checks can be eliminated.

\item The assignment {\bf \_cfg\_counter:=0} is redundant (line 24).

\item Conditions {\bf str[s+2]='a'} and {\bf str[s]='a'} are always false
      (lines 78 and 86, respectively) because two different constants are equal.
      So, the code of {\bf THEN}--branches is dead.

\item The conditions at the lines 63, 68, 74, 82, 104, 109 are false, too.
      However, automatic detection of these properties are not as easily as the previous.

\end{itemize}

Using this semantic information, it is possible to build a new
program functionally equivalent to {\bf Match("ababb",str)}. In
text of the program given below the underlined code can be
eliminated.

\vskip\baselineskip

\fontsize{8}{10}\selectfont
{\setcounter{numline}{0}
\begin{tabular}{ll}                                                          \vspace*{-1pt}
MODULE Match;                               &\\
\vspace*{-1pt} \Z  FROM FIO IMPORT
File,Open,ReadChar,\lefteqn{\mbox{WriteInt,stdout;}}&\\
\vspace*{-1pt}

\Z  VAR  \_cfg\_counter : CARDINAL; str\_file : File;        &\\
\vspace*{-1pt}

\Z  TYPE \_TYPE354a04 = [0..20];            &\\
\vspace*{-1pt}

\Z  TYPE \_TYPE355004 = ARRAY \_TYPE354a04 \lefteqn{\mbox{OF
CHAR;}}&\\      \vspace*{-1pt}

\Z  VAR str : \_TYPE355004; ls,s : \_TYPE354a04;
&\\          \vspace*{-1pt}
BEGIN                                       &\\\NM
\Z  str\_file := Open("target.dat");        &\\\NM
\Z  ls := 0;                                &\\\NM
\Z  LOOP                                    &\\\NM
\Z\Z  str[ls] := ReadChar(str\_file);       &\\\NM
\Z\Z  IF (str[ls]='\#') THEN                &$(*str[ls]='\!\!\#'*)$\\\NM
\Z\Z\Z  EXIT                                &\\\NM
\Z\Z  ELSE                                  &$(*str[ls]\neq'\!\!\#'*)$\\\NM
\Z\Z\Z  INC(ls)                             &\\\NM
\Z\Z  END                                   &\\\NM
\Z  END;                                    &$(*str[ls]='\!\!\#'*)$\\\NM
\Z  s := 0;                                 &\\\NM
\Z  \_cfg\_counter := 0;                    &$(*s=\_cfg\_counter=0*)$\\
\end{tabular}

\begin{tabular}{ll}                         \NM
\Z  LOOP                                    &\\\NM
\Z\Z    CASE \_cfg\_counter OF              &\\\NM
\Z\Z    $|$ 0 :                             &$(*\_cfg\_counter=0*)$\\\NM
\Z\Z\Z    IF ((s+0)$\geq$ls) THEN           &$(*s\geq ls*)$\\\NM
\Z\Z\Z\Z    WriteInt(stdout,(-1),0);        &\\\NM
\Z\Z\Z\Z    EXIT                            &\\\NM
\Z\Z\Z    END;                              &$(*0<ls,s<ls*)$\\\NM
\Z\Z\Z    IF (str[(s+0)]='a') THEN          &$(*str[s]='\!\!a',0<ls,s<ls*)$\\\NM
\Z\Z\Z\Z    \_cfg\_counter := 1             &\\\NM
\Z\Z\Z    ELSE                              &$(*0<ls,s<ls,str[0]\neq'\!\!a'*)$\\\NM
\Z\Z\Z\Z    INC(s);                         &$(*\_cfg\_counter=0*)$\\\NM
\Z\Z\Z\Z    \underline{\_cfg\_counter := 0} &\\\NM
\Z\Z\Z    END                               &$(*0<ls,s<ls*)$\\\NM
\Z\Z    $|$ 1 :                             &$(*\_cfg\_counter=1*)$\\\NM
\Z\Z\Z    IF ((s+1)$\geq$ls) THEN           &$(*s+1\geq ls*)$\\\NM
\Z\Z\Z\Z    WriteInt(stdout,(-1),0);        &\\\NM
\Z\Z\Z\Z    EXIT                            &\\\NM
\Z\Z\Z    END;                              &$(*s+1<ls*)$\\\NM
\Z\Z\Z    IF (str[(s+1)]='b') THEN          &$(*str[s+1]='\!\!b'*)$\\\NM
\Z\Z\Z\Z    \_cfg\_counter := 2             &\\\NM
\Z\Z\Z    ELSE                              &$(*str[s+1]\neq'\!\!b'*)$\\\NM
\Z\Z\Z\Z    INC(s);                         &$(*str[s]\neq'\!\!b',s<ls*)$\\\NM
\Z\Z\Z\Z    \_cfg\_counter := 0             &\\\NM
\Z\Z\Z    END                               &\\\NM
\Z\Z    $|$ 2 :                           &$(*\_cfg\_counter=2*)$\\\NM
\Z\Z\Z    IF ((s+2)$\geq$ls) THEN         &$(*s+2\geq ls*)$\\\NM
\Z\Z\Z\Z    WriteInt(stdout,(-1),0);      &\\\NM
\Z\Z\Z\Z    EXIT                          &\\\NM
\Z\Z\Z    END;                            &$(*s+2<ls*)$\\\NM
\Z\Z\Z    IF (str[(s+2)]='a') THEN        &$(*str[s+2]='\!\!a'*)$\\\NM
\Z\Z\Z\Z    \_cfg\_counter := 3           &\\\NM
\Z\Z\Z    ELSE                            &$(*str[s+2]\neq'\!\!a'*)$\\\NM
\Z\Z\Z\Z    INC(s);                       &$(*str[s+1]\neq'\!\!a',s+1<ls*)$\\\NM
\Z\Z\Z\Z    \_cfg\_counter := 4           &\\\NM
\Z\Z\Z    END                             &\\\NM
\Z\Z    $|$ 3 :                           &$(*\_cfg\_counter=3*)$\\\NM
\Z\Z\Z    IF ((s+3)$\geq$ls) THEN         &$(*s+3\geq ls*)$\\\NM
\Z\Z\Z\Z    WriteInt(stdout,(-1),0);      &\\\NM
\Z\Z\Z\Z    EXIT                          &\\\NM
\Z\Z\Z    END;                            &$(*s+3<ls*)$\\\NM
\Z\Z\Z    IF (str[(s+3)]='b') THEN        &$(*str[s+3]='\!\!b'*)$\\\NM
\Z\Z\Z\Z    IF ((s+4)$\geq$ls) THEN       &$(*s+4\geq ls,s+3<ls*)$\\\NM
\Z\Z\Z\Z\Z    WriteInt(stdout,(-1),0);    &\\\NM
\Z\Z\Z\Z\Z    EXIT                        &\\\NM
\Z\Z\Z\Z    END;                          &$(*s+4<ls*)$\\\NM
\Z\Z\Z\Z    IF (str[(s+4)]='b') THEN      &$(*str[s+3]=str[s+4]='\!\!b'*)$\\\NM
\Z\Z\Z\Z\Z    WriteInt(stdout,s,0);       &\\\NM
\Z\Z\Z\Z\Z    EXIT                        &\\\NM
\Z\Z\Z\Z    ELSE                          &$(*str[s+3]='\!\!b',str[s+4]\neq'\!\!b',s+4<ls*)$\\\NM
\Z\Z\Z\Z\Z    INC(s);                     &$(*str[s+2]='\!\!b',str[s+3]\neq'\!\!b',s+3<ls*)$\\\NM
\Z\Z\Z\Z\Z    \underline{IF ((s+0)$\geq$ls) THEN}     &$(*s\geq ls,s+3<ls*)$\\\NM
\Z\Z\Z\Z\Z\Z    \underline{WriteInt(stdout,(-1),0);}  &\\\NM
\Z\Z\Z\Z\Z\Z    \underline{EXIT}          &\\\NM
\Z\Z\Z\Z\Z    \underline{END;}            &$(*str[s+2]='\!\!b',str[s+3]\neq'\!\!b',s+3<ls*)$\\\NM
\Z\Z\Z\Z\Z    IF (str[(s+0)]='a') THEN    &$(*str[s]='\!\!a',str[s+2]='\!\!b',str[s+3]\neq'\!\!b',$\\
                                          &$\opn  s+3<ls*)$\\\NM
\Z\Z\Z\Z\Z\Z    \underline{IF ((s+1)$\geq$ls) THEN}   &$(*s+1\geq ls,s+3<ls*)$\\\NM
\Z\Z\Z\Z\Z\Z\Z    \underline{WriteInt(stdout,(-1),0);}&\\\NM
\Z\Z\Z\Z\Z\Z\Z    \underline{EXIT}        &\\\NM
\Z\Z\Z\Z\Z\Z    \underline{END;}          &$(*\mbox{the same as at line 67}*)$\\
\end{tabular}

\begin{tabular}{ll}                        \NM
\Z\Z\Z\Z\Z\Z    IF (str[(s+1)]='b') THEN  &$(*str[s]='\!\!a',str[s+1]='\!\!b',str[s+2]='\!\!b',$\\\NM
                                          &$\opn  str[s+3]\neq'\!\!b',s+3<ls*)$\\\NM
\Z\Z\Z\Z\Z\Z\Z    \underline{IF ((s+2)$\geq$ls) THEN} &$(*s+2\geq ls,s+3<ls*)$\\\NM
\Z\Z\Z\Z\Z\Z\Z\Z    \underline{WriteInt(stdout,(-1),0);}&\\\NM
\Z\Z\Z\Z\Z\Z\Z\Z    \underline{EXIT}      &\\\NM
\Z\Z\Z\Z\Z\Z\Z    \underline{END;}        &$(*str[s]='\!\!a',str[s+1]='\!\!b',str[s+2]='\!\!b',$\\
                                          &$\opn  str[s+3]\neq'\!\!b',s+3<ls*)$\\\NM
\Z\Z\Z\Z\Z\Z\Z    \underline{IF (str[(s+2)]='a') THEN}&$(*\mbox{inaccessible point}*)$\\\NM
\Z\Z\Z\Z\Z\Z\Z\Z    \underline{\_cfg\_counter := 3}   &\\\NM
\Z\Z\Z\Z\Z\Z\Z    \underline{ELSE}        &$(*str[s]='\!\!a',str[s+1]='\!\!b',str[s+2]='\!\!b',$\\
                                          &$\opn  str[s+3]\neq'\!\!b',s+3<ls*)$\\\NM
\Z\Z\Z\Z\Z\Z\Z\Z    INC(s);               &$(*str[s-1]='\!\!a',str[s]='\!\!b',str[s+1]='\!\!b',$\\
                                          &$\opn  str[s+2]\neq'\!\!b',s+2<ls*)$\\\NM
\Z\Z\Z\Z\Z\Z\Z\Z    \underline{IF ((s+0)$\geq$ls) THEN}&$(*s\geq ls,s+2<ls*)$\\\NM
\Z\Z\Z\Z\Z\Z\Z\Z\Z    \underline{WriteInt(stdout,(-1),0);}&\\\NM
\Z\Z\Z\Z\Z\Z\Z\Z\Z    \underline{EXIT}    &\\\NM
\Z\Z\Z\Z\Z\Z\Z\Z    \underline{END;}      &$(*str[s-1]='\!\!a',str[s]='\!\!b',str[s+1]='\!\!b',$\\
                                          &$\opn  str[s+2]\neq'\!\!b',s+2<ls*)$\\\NM
\Z\Z\Z\Z\Z\Z\Z\Z    \underline{IF (str[(s+0)]='a') THEN}&$(*\mbox{inaccessible point}*)$\\\NM
\Z\Z\Z\Z\Z\Z\Z\Z\Z    \underline{\_cfg\_counter:=14}  &\\\NM
\Z\Z\Z\Z\Z\Z\Z\Z    \underline{ELSE}      &$(*str[s-1]='\!\!a',str[s]='\!\!b',str[s+1]='\!\!b',$\\
                                          &$\opn  str[s+2]\neq'\!\!b',s+2<ls*)$\\\NM
\Z\Z\Z\Z\Z\Z\Z\Z\Z    INC(s);             &$(*str[s-2]='\!\!a',str[s-1]='\!\!b',str[s]='\!\!b',$\\
                                          &$\opn  str[s+1]\neq'\!\!b',s+1<ls*)$\\\NM
\Z\Z\Z\Z\Z\Z\Z\Z\Z    \_cfg\_counter:=4   &\\\NM
\Z\Z\Z\Z\Z\Z\Z\Z    \underline{END}       &\\\NM
\Z\Z\Z\Z\Z\Z\Z    \underline{END}         &$(*str[s-2]='\!\!a',str[s-1]='\!\!b',str[s]='\!\!b',$\\
                                          &$\opn  str[s+1]\neq'\!\!b',s+1<ls*)$\\
\NM
\Z\Z\Z\Z\Z\Z    ELSE                   &$(*str[s]='\!\!a',str[s+1]\neq'\!\!b',str[s+2]='\!\!b',$\\
                                       &$\opn  str[s+3]\neq'\!\!b',s+3<ls*)$\\\NM
\Z\Z\Z\Z\Z\Z\Z    INC(s);              &$(*str[s-1]='\!\!a',str[s]\neq'\!\!b',str[s+1]='\!\!b',$\\
                                       &$\opn  str[s+2]\neq'\!\!b',s+2<ls*)$\\\NM
\Z\Z\Z\Z\Z\Z\Z    \_cfg\_counter := 12 &\\\NM
\Z\Z\Z\Z\Z\Z    END                    &\\\NM
\Z\Z\Z\Z\Z    ELSE                     &$(*str[s]='\!\!a',str[s+2]='\!\!b',str[s+3]\neq'\!\!b',$\\
                                       &$\opn  s+3<ls*)$\\\NM
\Z\Z\Z\Z\Z\Z    INC(s);                &$(*str[s-1]='\!\!a',str[s+1]='\!\!b',str[s+2]\neq'\!\!b',$\\
                                       &$\opn  s+2<ls*)$\\\NM
\Z\Z\Z\Z\Z\Z    \_cfg\_counter := 12   &\\\NM
\Z\Z\Z\Z\Z    END                      &\\\NM
\Z\Z\Z\Z    END                        &\\\NM
\Z\Z\Z    ELSE                         &$(*str[s+3]\neq'\!\!b',s+3<ls*)$\\\NM
\Z\Z\Z\Z    INC(s);                    &$(*str[s+2]\neq'\!\!b',s+2<ls*)$\\\NM
\Z\Z\Z\Z    \underline{IF ((s+0)$\geq$ls) THEN}    &$(*s\geq ls,s+2<ls*)$\\\NM
\Z\Z\Z\Z\Z    \underline{WriteInt(stdout,(-1),0);} &\\\NM
\Z\Z\Z\Z\Z    \underline{EXIT}         &\\\NM
\Z\Z\Z\Z    \underline{END;}           &$(*str[s+2]\neq'\!\!b',s+2<ls*)$\\\NM
\Z\Z\Z\Z    IF (str[(s+0)]='a') THEN   &$(*str[s]='\!\!a',str[s+2]\neq'\!\!b',s+2<ls*)$\\\NM
\Z\Z\Z\Z\Z    \underline{IF ((s+1)$\geq$ls) THEN}  &$(*s+1\geq ls,s+2<ls*)$\\\NM
\Z\Z\Z\Z\Z\Z    \underline{WriteInt(stdout,(-1),0);}&\\\NM
\Z\Z\Z\Z\Z\Z    \underline{EXIT}       &\\\NM
\Z\Z\Z\Z\Z    \underline{END;}         &$(*str[s]='\!\!a',str[s+2]\neq'\!\!b',s+2<ls*)$\\
\end{tabular}

\begin{tabular}{ll}                    \NM
\Z\Z\Z\Z\Z    IF (str[(s+1)]='b') THEN &$(*str[s]='\!\!a',str[s+1]='\!\!b',str[s+2]\neq'\!\!b',$\\
                                       &$\opn  s+2<ls*)$\\\NM
\Z\Z\Z\Z\Z\Z    \_cfg\_counter := 2    &\\\NM
\Z\Z\Z\Z\Z    ELSE                     &$(*str[s]='\!\!a',str[s+2]\neq'\!\!b',s+2<ls*)$\\\NM
\Z\Z\Z\Z\Z\Z    INC(s);                &$(*str[s-1]='\!\!a',str[s+1]\neq'\!\!b',s+1<ls*)$\\\NM
\Z\Z\Z\Z\Z\Z    \_cfg\_counter := 10   &\\\NM
\Z\Z\Z\Z\Z    END                      &\\\NM
\Z\Z\Z\Z    ELSE                       &$(*str[s+2]\neq'\!\!b',s+2<ls*)$\\\NM
\Z\Z\Z\Z\Z    INC(s);                  &$(*str[s+1]\neq'\!\!b',s+1<ls*)$\\\NM
\Z\Z\Z\Z\Z    \_cfg\_counter := 10     &\\\NM
\Z\Z\Z\Z    END                        &\\\NM
\Z\Z\Z    END                          &\\\NM
\Z\Z    $|$ 4 :                        &$(*\_cfg\_counter=4*)$\\\NM
\Z\Z\Z    IF ((s+0)$\geq$ls) THEN      &$(*s\geq ls*)$\\\NM
\Z\Z\Z\Z    WriteInt(stdout,(-1),0);   &\\\NM
\Z\Z\Z\Z    EXIT                       &\\\NM
\Z\Z\Z    END;                         &$(*s<ls*)$\\\NM
\Z\Z\Z    IF (str[(s+0)]='a') THEN     &$(*str[s]='\!\!a',s<ls*)$\\\NM
\Z\Z\Z\Z    \_cfg\_counter := 1        &\\\NM
\Z\Z\Z    ELSE                         &$(*str[s]\neq'\!\!a',s<ls*)$\\\NM
\Z\Z\Z\Z    INC(s);                    &$(*str[s-1]\neq'\!\!a',s\leq ls*)$\\\NM
\Z\Z\Z\Z    \_cfg\_counter := 0        &\\\NM
\Z\Z\Z    END                          &\\\NM
\Z\Z    $|$ 10:                        &$(*\_cfg\_counter=10*)$\\\NM
\Z\Z\Z    IF ((s+0)$\geq$ls) THEN      &$(*s\geq ls*)$\\\NM
\Z\Z\Z\Z    WriteInt(stdout,(-1),0);   &\\\NM
\Z\Z\Z\Z    EXIT                       &\\\NM
\Z\Z\Z    END;                         &$(*s<ls*)$\\\NM
\Z\Z\Z    IF (str[(s+0)]='a') THEN     &$(*str[s]='\!\!a',s<ls*)$\\\NM
\Z\Z\Z\Z    \_cfg\_counter := 1        &\\\NM
\Z\Z\Z    ELSE                         &$(*str[s]\neq'\!\!a',s<ls*)$\\\NM
\Z\Z\Z\Z    INC(s);                    &$(*str[s-1]\neq'\!\!a',s\leq ls*)$\\\NM
\Z\Z\Z\Z    \_cfg\_counter := 0        &\\\NM
\Z\Z\Z    END                          &\\
\NM
\Z\Z $|$ 12:                           &$(*\_cfg\_counter=12*)$\\\NM
\Z\Z\Z    IF ((s+0)$\geq$ls) THEN      &$(*s\geq ls*)$\\\NM
\Z\Z\Z\Z    WriteInt(stdout,(-1),0);   &\\\NM
\Z\Z\Z\Z    EXIT                       &\\\NM
\Z\Z\Z    END;                         &$(*s<ls*)$\\\NM
\Z\Z\Z    IF (str[(s+0)]='a') THEN     &$(*str[s]='\!\!a',s<ls*)$\\\NM
\Z\Z\Z\Z    \_cfg\_counter := 14       &\\\NM
\Z\Z\Z    ELSE                         &$(*str[s]\neq'\!\!a',s<ls*)$\\\NM
\Z\Z\Z\Z    INC(s);                    &$(*str[s-1]\neq'\!\!a',s\leq ls*)$\\\NM
\Z\Z\Z\Z    \_cfg\_counter := 4        &\\\NM
\Z\Z\Z    END                          &\\\NM
\Z\Z $|$ 14:                           &$(*\_cfg\_counter=14*)$\\\NM
\Z\Z\Z    IF ((s+1)$\geq$ls) THEN      &$(*s+1\geq ls*)$\\\NM
\Z\Z\Z\Z    WriteInt(stdout,(-1),0);   &\\\NM
\Z\Z\Z\Z    EXIT                       &\\\NM
\Z\Z\Z    END;                         &$(*s+1<ls*)$\\\NM
\Z\Z\Z    IF (str[(s+1)]='b') THEN     &$(*str[s+1]='\!\!b',s+1<ls*)$\\\NM
\Z\Z\Z\Z    \_cfg\_counter := 2        &\\\NM
\Z\Z\Z    ELSE                         &$(*str[s+1]\neq'\!\!b',s+1<ls*)$\\\NM
\Z\Z\Z\Z    INC(s);                    &$(*str[s]\neq'\!\!b',s<ls*)$\\\NM
\Z\Z\Z\Z    \_cfg\_counter := 4        &\\\NM
\Z\Z\Z    END                          &\\\NM
\Z\Z  END                              &\\
\Z  END                                &\\
END Match.                             &\\
\end{tabular}

\end{document}